%% file: mateu_tkdprofiles.tex
\documentclass[useAMS,usenatbib]{mnras}

\usepackage[british]{babel}             
\usepackage{graphicx}

\defcitealias{Vivas2004}{V04}
\defcitealias{Layden1995}{L95}
\defcitealias{Mateu2012}{M12}
\defcitealias{Sesar2010}{S10}
\usepackage{times}
\usepackage{amsmath}
\usepackage{amssymb}
\usepackage{color,graphicx}
\usepackage{multirow}
\usepackage{booktabs}
\usepackage{bm}
\usepackage[flushleft]{threeparttable}
\usepackage{longtable}

\usepackage{hyperref}

\newcommand{\lsim}{\raisebox{-.5ex}{$\,\stackrel{\textstyle <}{\sim}\,$}}

\footnotesize{\tiny}

\newcommand{\rrab} {\mbox{RR\emph{ab}}}
\newcommand{\rrc} {\mbox{RR\emph{c}}}
\newcommand{\typeab} {\mbox{\emph{ab}}}
\newcommand{\typec} {\mbox{\emph{c}}}
\newcommand{\tkd} {{\mathrm{\scriptscriptstyle TkD}}}

\newcommand{\h} {{\mathrm{\scriptscriptstyle H}}}
\newcommand{\hel} {{\mathrm{hel}}}
\newcommand{\FeH} {[\mathrm{Fe}/\mathrm{H}]}
\newcommand{\alphaFe} {[\alpha/\mathrm{Fe}]}
\newcommand{\rrabkpc}{\rrab~kpc$^{-3}$}
\newcommand{\rrkpc}{RRLS~kpc$^{-3}$}
\newcommand{\nkpc}{\#~kpc$^{-3}$}
\newcommand{\Ron}{R_{\mathrm{on}}}
\newcommand{\hrf}{h_{\mathrm{rf}}}

\def\apj{ApJ}
\def\apjs{ApJS}
\def\apjl{ApJL}
\def\aj{AJ}
\def\mnras{MNRAS}
\def\aa{A\&A}

\def\araa{ARA\&A}
\def\aap{A\&A}
\def\pasp{PASP}

\title[Thick disc Profile with RRLS]{The Galactic thick disc density profile traced with RR Lyrae stars}
\author[Mateu \& Vivas]{Cecilia Mateu,$^{1}$\thanks{cmateu@cida.gob.ve, cecilia.mateu@gmail.com} A. Katherina Vivas,$^{2}$\thanks{kvivas@ctio.noao.edu} \\
$^{1}${{Centro de Investigaciones de Astronom\'{\i}a, AP 264, M\'erida 5101-A, Venezuela}}\\
$^{2}${{Cerro Tololo Inter-American Observatory, National Optical Astronomy
Observatory, Casilla 603, La Serena, Chile.}}\\
}

\begin{document}

\maketitle

\label{firstpage}

\begin{abstract}
We used a combination of public RR Lyrae star catalogs and a Bayesian methodology to derive robust structural parameters of the inner halo ($<25$~kpc) and thick disc of the Milky Way. RR Lyrae stars are an unequivocal tracer of old metal-poor populations, for which accurate distances and extinctions can be individually estimated and so, are a reliable independent means of tracing the population of the old high-$\alphaFe$ disc usually associated to the thick disc. In particular, the chosen RR Lyrae sample spans regions at low galactic latitude toward the anti-center direction, allowing to probe the outermost parts of the disc. Our results favour a thick disc with short scale height and short scale length, $h_z=0.65_{-0.05}^{+0.09}$~kpc, $h_R=2.1_{-0.25}^{+0.82}$~kpc, for a model in which the inner halo has a constant flattening of $q=0.90_{-0.03}^{+0.05}$ and a power law index of $n=-2.78_{-0.05}^{+0.05}$. Similar short scales for the thick disc are also found when considering an inner halo with flattening dependent on radius. We also explored a model in which the thick disc has a flare and, although this is only mildly constrained with our data, a flare onset in the inner $\sim11$~kpc is highly disfavoured. 

\end{abstract}


\section{INTRODUCTION}\label{s:intro}

Even now, over thirty years after its discovery, the Galactic thick disc remains a controversial topic. Whether or not it is a distinct component \citep{Bovy2011,Rix2013,Kawata2016} and how does its chemistry, spatial extent and kinematics relate to those of the thin disc, halo and bulge \citep[e.g.][]{Bensby2004,Bensby2007,Bensby2009,Reddy2006,RecioBlanco2014}, are current subjects of debate.  Nonetheless, regardless of whether or not it is a distinct component, from the standpoint of disc formation its importance is clear: the thick disc contains the \emph{oldest most metal-poor stars in the disc}, which hold information of the earliest processes that occurred during the formation of the Galactic disc. 

To study the properties of the Galactic disc based on stars selected by age is notoriously hard due to the difficulty of computing accurate ages for field stars, particularly for distant ones. One way to go around this problem is to use chemical abundances as a proxy for age. Different star formation histories produce well defined trends in the $\alphaFe-\FeH$ plane \citep{Matteucci2003}, so a broad component/age classification can be made by selecting broadly separated regions in this plane \citep[e.g.][]{Navarro2010,Kordopatis2015}.  Pushing this idea further, \citet{Bovy2012a} defines mono-abundance sub-populations (MAPs) as bins in the $\alphaFe-\FeH$ plane, taking advantage of the large number of stars with spectroscopic $\alphaFe$ and $\FeH$ abundances from the SEGUE survey. An advantage of the selection in chemical abundance space is that it is independent of kinematics and spatial considerations and, thus, kinematic and structural properties can be studied without the biases and correlations typically introduced when inferring kinematic properties from a spatially selected sample or viceversa \citep{Freeman2002,Navarro2010,Bovy2012a}. On the other hand, it has been pointed out that MAPs do not translate univocally into single-age distributions \citep{Minchev2017} and this approach requires medium to high-resolution spectroscopy, hence it will not be immediately applicable to \emph{Gaia}, which will provide astrometric and photometric data for stars down to magnitude $G=20$, but only
low-resolution spectroscopy for the brightest stars ($G\leqslant15$).

Another way to go about selecting stars by age is to use knowledge on stellar populations to choose types of stars known for tracing well defined age ranges \citep{Grebel1999}: e.g. classical or type I Cepheids, which trace young populations \citep{Sandage2006}; long period Miras and TP-AGB stars, found in intermediate age populations \citep{Catchpole2016,Feast2000}, or RR Lyrae stars (RRLSs) which trace old and metal-poor populations \citep[$\geqslant10$ Gyr and $\FeH\lesssim-0.5$, e.g.][]{Smith1995}. Many of these stellar tracers, in particular pulsating stars such as RRLSs, Cepheids and Miras, can be confidently identified photometrically, so this approach has the advantage of also avoiding kinematic and structural biases, without requiring spectroscopic data. 

In this work we use RRLSs which are outstanding tracers to study the structure of the old and metal-poor disc, as they are unequivocal tracers of populations with ages $\geqslant 10$~Gyr. RRLSs are well known for being excellent standard candles, for which accurate distances can be estimated with errors $\sim$5--7$\%$ from optical data alone \citep{Vivas2006,Sesar2010,Mateu2012}, or even down to $\sim$1--2$\%$ when folding in infrared data and/or metallicity information \citep{Neeley2017,Madore2012,Sesar2017b}. RRLSs are also colour standards: during a particular phase range at minimum light RRLSs have the same temperature and, therefore, the same intrinsic colour \citep{Sturch1966,Guldenschuh2005,Kunder2010,Vivas2017}. This is a particularly important property for studies of RRLSs at low latitude, where the extinction is not only high but also highly variable, as it allows to compute individual extinctions for each star based on its colour time series. Also, since they are present only in stellar populations that are simultaneously old and metal-poor, RRLSs are found in the Galactic halo, bulge and thick disc, but no significant numbers have been observed to have thin disc kinematics \citep{Martin1998}.
 
RRLSs have been used as tracers of the thick disc structure in three previous studies: \citet{Layden1995} and \citet{Amrose2001} estimated the thick disc scale height and \citet{Kinemuchi2006}, who estimated the scale height and also analysed the Oosterhoff properties of thick disc RRLSs. However none of these authors, nor any other study to date, have used RRLSs to estimate the scale length, a property of the thick disc which remains highly uncertain.  

Recent works by \citet{Bovy2011,Bovy2012,Bovy2016}, using RC and RGB stars as tracers, and \citet{Robin2014}, via stellar population synthesis, favour a relatively short scale height -- scale length combination, around $h_z\sim0.5-0.6$ and $h_R\sim2.2$~kpc respectively, in contrast to most previous works which favoured larger values for both parameters around $h_z\sim0.9$ and $h_R\sim3.5$~kpc; illustrating the large scatter found in the literature for thick disc parameters. Also, the limited knowledge of the thick disc structure regarding the possible existence of a flare, warp or even suggested disc oscillations \citep{LopezCorredoira2002,Momany2006,Xue2015,Newberg2017}, has been at the centre of the controversy about the nature of substructures at low latitude such as the Monoceros Ring, the Canis Major (CMa) overdensity and the Tri-And cloud. Here too, RRLSs have played a central role as \citet{Mateu2009} and \citet{PriceWhelan2015} have argued CMa and Tri-And to be more likely disc related perturbations than accreted tidal debris, by comparing the observed RRLS content to the expectations from a dwarf galaxy or the disc's population.

Here we use a combination of public RRLS surveys to provide the first estimation of the scale length, height and local density normalization of Galactic disc RRLSs. We will also explore a flared thick disc model and present the first estimation of the flare parameters derived with RRLSs. Our analysis follows the same methodology as \citet{Bovy2012b} to infer the density profile parameters, but with an entirely different tracer that, as mentioned, is reliably identified with an old and metal-poor population and provides precise individual distances. Using this methodology coupled with RRLSs has the considerable advantage of using solely photometric information, so it will be readily applicable to surveys like Pan-STARRS and \emph{Gaia} and LSST coming in the near future.

The structure of the paper is as follows: in Section~\ref{s:dens_prof} we will describe the methodology used to fit the density profile models; in Section~\ref{s:rrl_samples} we described the RRLS catalogues used; in Sections~\ref{s:res_halo_normal_tkd} and \ref{s:res_flared_disk} we describe our results for the halo and normal thick disc and for the flared thick disc respectively; in Sec.\ref{s:comp_prev_aut} we compare our results with those of previous studies and  in Section~\ref{s:conclusions} we summarise our conclusions.

\section{Fitting Galactic Density Profiles}\label{s:dens_prof}

\subsection{Bayesian Fitting}\label{s:bayesmodel}

To fit the density profile models by comparing them with the observations of the position ${\bf r}^{RRLS}$ of RRLSs, we use the Bayesian methodology described in \citet{Lombardi2013}, used also in \citet{Bovy2012b}.  This consists in using a Bayesian forward model to compare the observed distribution of stars with the expected distribution for a model with density profile $\rho(h_z,h_R,C_\tkd,n,q,C_\h)$, and expressing the likelihood as that given by a Poisson process:

\begin{equation}\label{e:lnL}
 \ln L = \sum_{i=1}^{N^{RRL}_{obs}} \ln\rho({\bf \theta},{\bf r}_i^{RRLS}) - N^{RRL}_{model}({\bf \theta})
\end{equation}

\noindent
where the sum is performed over the $N^{RRL}_{obs}$ position vectors ${\bf r}^{RRLS}$ of the \emph{observed} RRLSs and $N^{RRL}_{\rm model}({\bf \theta})$ is the total number of RRLSs predicted by the model with parameters ${\bf \theta}=(h_z,h_R,C_\tkd,n,q,C_\h)$ inside the surveyed volume and accounting for the selection function, given by

\begin{equation}\label{e:nrr_model}
N^{RRL}_{\rm model}({\bf \theta}) = \int_{V_S} dV \rho({\bf \theta},{\bf r}) I({\bf r})
\end{equation}

Here, $I({\bf r})$ describes the selection function, i.e. the completeness of the survey as a function of position ${\bf r}$, and $V_S$ denotes the survey volume over which the selection function is valid. This form for the likelihood, in terms of $N^{RRL}_{\rm model}$, allows us to express it easily in observable quantities, as $N^{RRL}_{\rm model}$, $V_S$ and $I({\bf r})$ can be expressed in a simple manner as

\begin{multline}
N^{RRL}_{\rm model}({\bf \theta}) = \int_{\alpha_o}^{\alpha_f}\!\int_{\delta_o}^{\delta_f}\!\int_{R_\hel^o}^{R_\hel^f}\!  
\rho({\bf \theta},{\bf r}(\alpha,\delta,R_\hel)) I({\alpha,\delta,R_\hel}) \\
 R_\hel^2\cos{\delta} \,d\alpha d\delta dR_\hel
\end{multline}

In this equation we have expanded the volume integral and volume element to their explicit forms in equatorial coordinates $(\alpha,\delta)$ and heliocentric distance $R_\hel$, in which they are easily expressed. The detailed volume limits and selection function for each survey are discussed in Section~\ref{s:survey_volume}. Note also that the RRLS selection function is quite simple as it does not depend on stellar parameters such as colour or metallicity as it does in \citet{Bovy2012b}. 
  
From Bayes' Theorem, the (log) posterior probability density function (PDF) is simply

\begin{equation}
\ln P(\theta,C_\h | \{{\bf r}_i^{RRLS}\}) = \sum_{j}^{} \ln L_j + \ln P(\mathbf{\theta})
\end{equation}

where $P({\bf \theta})$ is the prior PDF on the density profile model parameters and $\ln L_j$ is the log-likelihood for each separate RRLS survey, given by Equations~\ref{e:lnL} and \ref{e:nrr_model}. 

The density profile models are described in the next section. For all model parameters we have assumed linear uniform priors, with limits summarised in Table~\ref{t:priorlims}. 
The choice of minimum and maximum values for the model parameters was guided by results from previous works, which we summarise in Section~\ref{s:comp_prev_aut}, and from initial test evaluations of the posterior PDF which, as discussed in Section~\ref{s:nrr_computation}, was computed by direct evaluation.

\begin{table}
\begin{center}
\tabcolsep=0.1cm
\begin{small}
\begin{tabular}{cccccl}
\hline
Parameter & Units & Min & Max & Spacing & Meaning \\
\hline
$n$         & $\cdots$ & -3.3& -2.1  & 0.04 & halo power law exponent \\
$q$         & $\cdots$ & 0.5 & 1.5   & 0.05 & halo flattening \\
$c_\h$    & \nkpc  & 1.6     & 8.0   &   0.4 &  halo local number density \\
$h_z$     & kpc    & 0.3     & 1.2    & 0.05 & thick disc scale height \\
$h_R$    & kpc    & 1.5     & 5.0    & 0.1   &  thick disc scale length \\
$C_\tkd$& \nkpc  & 6.0     & 30.0 & 1.25 &  thick disc local number density \\
$h_{rf}$  & kpc    & 0.2     & 4.0   &  0.2  & flare scale length \\
$R_{on}$& kpc    & 6.0     & 18.0  & 0.2  & flare onset radius \\
\hline
\end{tabular}\end{small}
\caption{Uniform prior limits assumed for density profile parameters.}
\label{t:priorlims}
\end{center}
\end{table}

\subsection{Density Profile Models $\rho({\bf \theta},{\bf r}$) }\label{s:model_rhos}

Since we will not do an a priori assignment of individual RRLSs~to each Galactic component, the total density profile is expressed as the sum of the halo and thick disc density profiles $\rho_\h({\bf r})$ and $\rho_\tkd({\bf r}$) respectively, as

\begin{equation}\label{e:rho_total}
 \rho({\bf r}) = \rho_\h({\bf r}) + \rho_\tkd({\bf r})
\end{equation}

\noindent


In Eq.~\ref{e:rho_total} we have assumed there is no contribution of either the thin disc or the bulge to the density profile. The possible contamination of these two components in our observational sample is further discussed in Section~\ref{s:contamination}.

\subsubsection{Halo}\label{s:halo_models}

For the halo we use a standard power law density profile:

\begin{equation} \label{e:rho_halo} 
 \rho_\h (R,z) = \frac{C_\h}{R_\odot^n} \Big[ R^2 + \Big(\frac{z}{q}\Big)^2 \Big]^{n/2} 
\end{equation}

\noindent
where $R$ and $z$ are respectively the cylindrical radius and height above the disc and $q$ is the flattening. Two different models are considered for the halo flattening law: \emph{halo Model A}, with a constant flattening $q$ as a free parameter; and \emph{halo Model B}, in which $q$ has a fixed dependency with radius according to \citet{Preston1991}, as follows:

\begin{equation}\label{e:qpreston}
q = \left\{ \begin{array}{lll} 
q_\circ + (1-q_\circ)(a/a_u) & \mathrm{if} & a<a_u \\
1 &  \mathrm{if} & a \geqslant a_u \\
\end{array} \right. 
 \end{equation}
 
 with $a^2 = R^2 + z^2q^{-2}$. We have assumed $q_\circ=0.5$ and $a_u=20$ kpc following  \citet{Preston1991} and \citet{Vivas2006}. 
 
 Most recent works on number density profiles of halo RRLSs are focused on the outer halo, in which a constant flattening (halo Model A) seems appropriate and enough to model the data \citep[e.g.][]{Sesar2007,Zinn2014}. However, a variable flattening in the inner halo was proposed by \citet{Preston1991} to model earlier ``in-situ" observations by \citet{Kinman1965} of RRLSs along lines of sights close to the Center of the Galaxy ($l = 11\degr, b = 30\degr$). These early data show a clear flattening of the halo compared with the number density profile obtained along other directions in the Galaxy. \citet{Vivas2006} found that halo Model B was needed to explain also the observed density profile of QUEST RRLSs in directions close to $l=0\degr$. Although the motivation for this model is completely empirical, there has been accumulating evidence over the last decades that the halo may be the result of both in-situ formation and accretion, with each mechanism dominating the inner and outer halo respectively \citep{Zinn1993,Carollo2007,Carollo2010,Kinman2012}. The different flattening in the inner halo suggested in halo Model B may be a consequence of the different mechanism of formation of each component. Since our dataset includes the QUEST data \citep{Vivas2004}, which has lines of sights toward $l=0\degr$, it seemed appropriate to include this model too in our analysis.

Several authors \citep{Saha1985,Carollo2010,Watkins2009, Deason2011,Zinn2014} have also suggested a possible change of the slope $n$ for the outer halo ($>20-27$~kpc). To avoid dealing a more complicated halo model,  we have restricted our analysis to RRLSs in the inner halo, at a Galactocentric distance $<$25 kpc (see also Sec.~\ref{s:survey_volume}).
  
\subsubsection{Thick disc} \label{s:normal_tkd_models}

For the standard thick disc we use a double exponential for the density profile:

\begin{equation}\label{e:rho_tkd} 
\rho_\tkd (R,z) =C_{\tkd} e^{-\tfrac{R-R_\odot}{h_R}} e^{-\tfrac{|z|}{h_z}} 
\end{equation}

For the flared model of the thick disc we use the double exponential model of Eq.~\ref{e:rho_tkd} and express the scale height as a function of radius $h_z(R)$  following \citet{Hammersley2011} and \citet{Conn2012}:

\begin{equation}\label{e:flare}
h_z(R) = \left\{ \begin{array}{lll} 
h_z^\odot &  \mathrm{if} & R \leqslant R_{on} \\
h_z^\odot e^{\tfrac{R-R_{on}}{h_{rf}}} & \mathrm{if} & R > R_{on} \\
\end{array} \right. 
 \end{equation}
 
Thus, $h_z(R)$ is constant ($h_z^\odot$) up to some flare onset radius $R_{on}$, after which it increases exponentially with a characteristic scale length $h_{rf}$.
 
\subsection{Computation of $N^{RRL}_{model}(\mathbf{\theta})$}\label{s:nrr_computation}

In the calculation of the likelihood of Eq.~\ref{e:lnL}, the only computationally intensive part is the calculation of $N^{RRL}_{model}$, the number of RRLSs predicted by the model for each given set of parameters. However, $N^{RRL}_{model}$ can be split into two independent calculations, separating the halo and thick disc contributions. Thus, $N^{RRL}_{\mathrm{H}}$ and $N^{RRL}_{\mathrm{TkD}}$ can be computed independently and then added together. For the relatively low number of parameters in our problem, this means we can compute the posterior PDF by direct evaluation in a dense N-point grid for each component (halo and thick disc), which take only $2N$ expensive computations (rather than $N^2$) that are then combined by simple addition to produce the full posterior PPDF. The computation by direct evaluation has the clear advantage of giving a straight-forward measurement of the posterior, without the worries related to Markov Chain Monte Carlo sampling like assuring proper burn-in and convergence and tempering to deal with multi-modality \citep[e.g.][]{ForemanMackey2013,Sharma2017}.

\section{The RRLS catalogues}\label{s:rrl_samples}

We used four publicly available RRLS catalogues in the literature: the QUEST RRLS catalogue, which combines the low and high galactic latitude samples from  \citet{Mateu2012} and \citet{Vivas2004}; the \citet{Layden1995} solar neighbourhood catalogue; and the \citet{Sesar2010} SDSS stripe 82 catalogue of intermediate and high galactic latitude stars. 

The selection of these catalogs was based on the need to have ample coverage on galactic latitude, specially at low $b$ values, while avoiding contribution by bulge stars. To understand the completeness of each catalog was another key aspect in the selection of the RRLS catalogues since it is required in Equation~\ref{e:rho_total}.

\begin{figure*}
\begin{center}
 \includegraphics[width=\columnwidth]{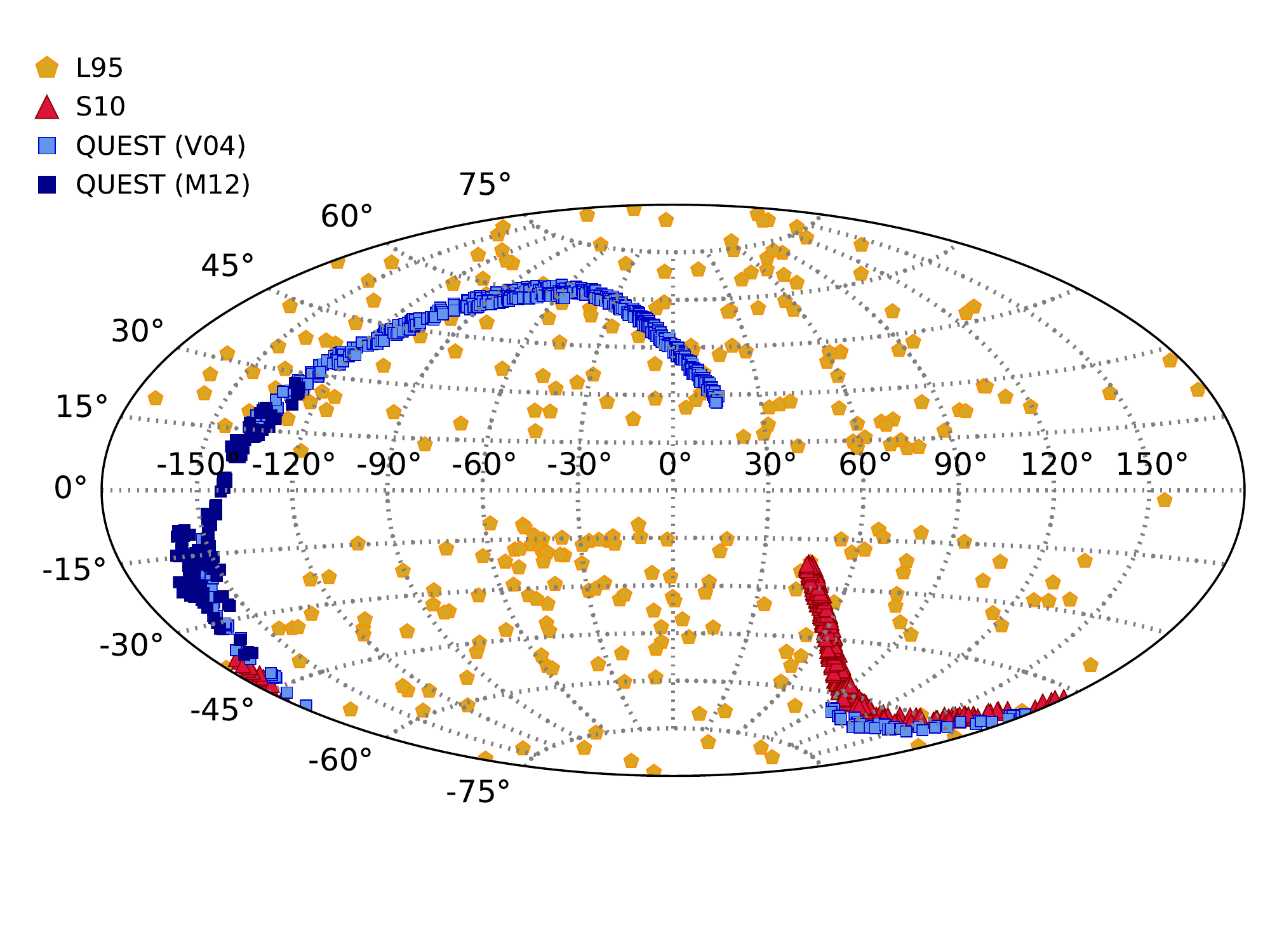}
 \includegraphics[width=\columnwidth]{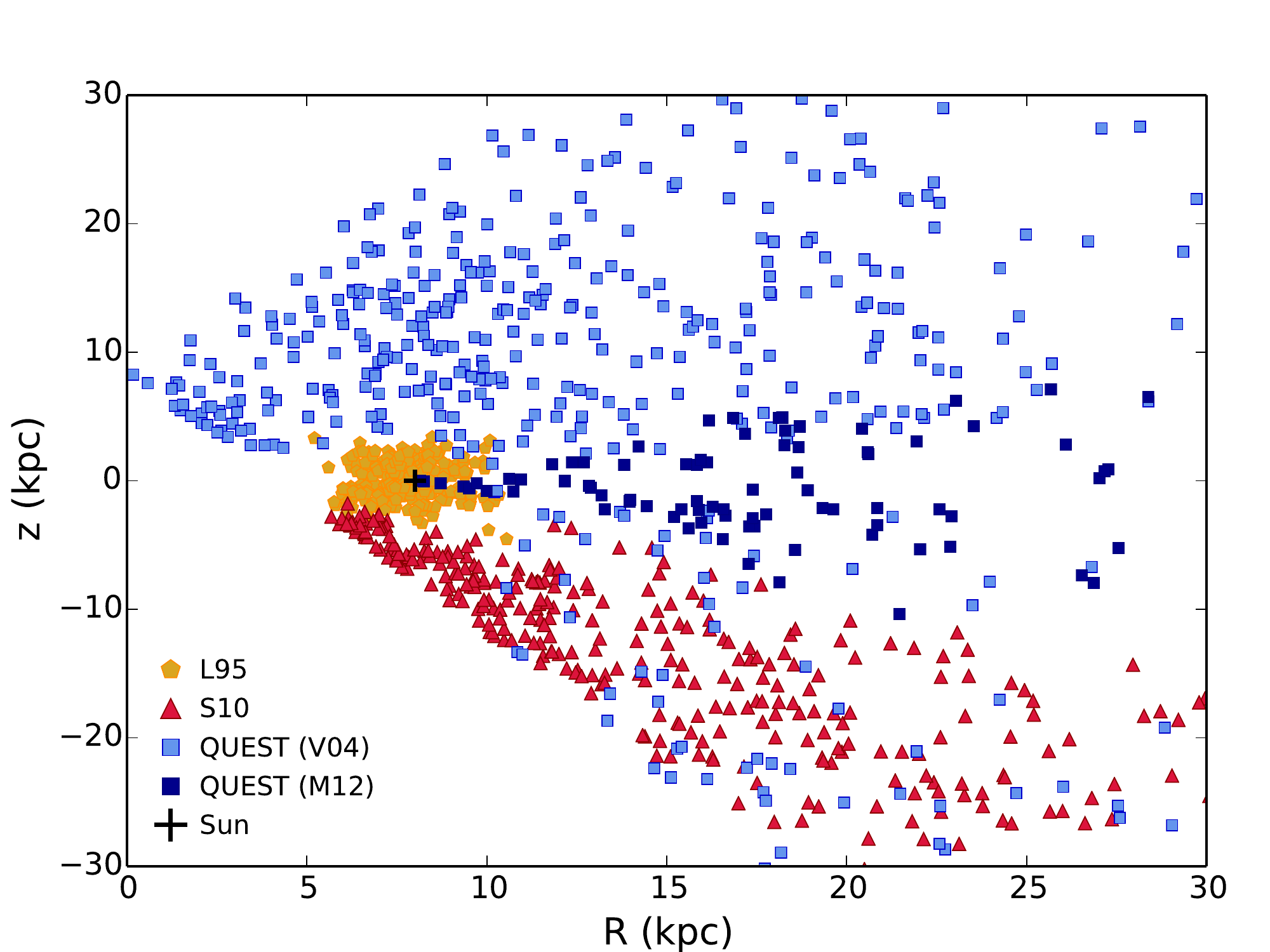} 
\caption{Footprints for the RRLSs surveys used in this work: L95 (pentagons), S10 (triangles), QUEST V04 (light filled squares) and QUEST M12 (dark filled squares). \emph{Left:} Aitoff projection map in galactic coordinates. \emph{Right:} Cylindric plot $z$ versus $R$. This plot is limited to $|z|<30$ kpc and $R<30$ kpc in order to emphasize the coverage of the samples around the Galactic disc, since the QUEST and S10 samples extend towards much larger distances and heights above the plane.}
\label{f:survey_coverage_aitoff}
\end{center}
\end{figure*}

Figure \ref{f:survey_coverage_aitoff} illustrates the coverage of the three surveys in an Aitoff map in galactic coordinates (\emph{left}) and in cylindric coordinates $z$ versus $R$ (\emph{right}). These plots show the combined RRLS catalogues span a large range of Galactic radii $6\lesssim R(\mathrm{kpc})\lesssim16$ at low height above the disc $|z|<5$~kpc, crucial for the derivation of thick disc parameters, while at the same time reaching up to $R>25$~kpc away from the disc, which allows us to properly estimate the halo contribution simultaneously. In the following sections we will describe the samples used, as well as the characterisation of the survey volumes and completeness.

\subsection{The QUEST RRLSs sample}

In the QUEST sample we have combined the RRLS catalogues from \citet{Vivas2004} and \citet{Mateu2012}, both of which were produced from drift-scan optical multi-epoch VRI observations obtained with the QUEST-I camera and the J\"urgen Stock telescope, at the Llano del Hato National Observatory of Venezuela. These two catalogues complement one another in galactic latitude: the V04 catalogue spans intermediate to high latitudes ($|b|\geq20\degr$), while the M12 catalogue extends across the Galactic Plane ($|b|<25\degr$) in the anti-center direction. 
 
In total, this QUEST catalogue has 847 RRLSs (637 of type \typeab~and 210 of \typec) and spans an area of $\sim850$ deg$^2$, extending in right ascension from $60\degr$ up to $210\degr$ in an equatorial stripe with varying declination coverage in the range $|\delta|\leq 6\fdg5$, and spanning a large heliocentric distance range $4<R_{\hel}({\rm kpc})<60$ (see Fig.~\ref{f:survey_coverage_aitoff}).

Thus, the combined QUEST catalogue provides a large radial coverage spanning from $10\lesssim R {\rm(kpc)}\lesssim 15$ at low height above the disc $|Z|<3$~kpc (see Figure~\ref{f:survey_coverage_aitoff}, right), a key point necessary to constrain the radial profile of the thick disc.

\subsection{The Layden et al. 1995 RRLSs sample}

The RRLS catalogue from \citet{Layden1995} (hereafter L95) is a compilation from the \textit{General Catalogue of Variable Stars}  \citep[GCVS4]{Kholopov1987}, which was combined with follow-up spectroscopy and photometry from \citet{Layden1994}, from which radial velocities, spectroscopic metallicities and distances were computed for $302$ \rrab~stars. The sample covers almost the entire sky, with a zone of avoidance at $|b|<10\degr$ (see Figure \ref{f:survey_coverage_aitoff}), and heliocentric distances in the range $0.1 \leqslant R_\hel ({\rm kpc}) \leqslant 2.5$. 
This catalogue is crucial to extend the radial coverage of the QUEST sample close to the Galactic Plane at Galactic radii $6\lesssim R \mathrm{(kpc)}\lesssim 10$ around the Sun.

\subsection{The Sesar et al. 2010 RRLSs sample}

This catalogue comprises 483 RRLSs (366 of type \typeab~and 117 of \typec) identified by \citet{Sesar2010} from SDSS-II stripe 82 observations, an equatorial strip from SDSS with multi-epoch {\it ugriz} observations in the right ascension range $300\degr<\alpha<60\degr$ and declination $|\delta|<1\fdg27$, spanning $\sim280$ deg$^2$ in the southern Galactic hemisphere. The S10 sample spans a large range in heliocentric distances  $4 \leqslant R_{\hel} ({\rm kpc}) \leqslant 150$, at intermediate to high latitudes $|b|>20\degr$ (see Figure \ref{f:survey_coverage_aitoff}). This catalogue increases radial and latitude coverage away from the disc, complementing the QUEST catalog at negative $z$, which helps to better constrain the halo profile (see Sec.~\ref{s:res_halo_normal_tkd}). 

\subsection{Distance Recomputation}

Since the heliocentric distances reported in the literature for each of the RRLS surveys were obtained using different methods, in order to homogenise the data we recomputed the distances to all RRLSs in the L95, S10 and V04 catalogues to match those in \citet{Mateu2012}, which uses the Period-Luminosity relation of \citet{Catelan2004}, and assuming \citet{Schlafly2011} extinctions for each star. The distances to the M12 RRLSs were originally computed using extinctions calculated individually for each star by comparing the observed $V-I$ and $V-R$ colours to the intrinsic colours from \citet{Guldenschuh2005}. The \citet{Guldenschuh2005} intrinsic colours were originally calibrated using \citet{Schlegel1998} extinctions, which have a small but systematic offset with respect to those of \citet{Schlafly2011}, so we corrected these and recomputed the M12 RRLSs distances accordingly.

The \citet{Catelan2004} Period-Luminosity relation requires the metallicity for each star. For the L95 RRLSs we used the spectroscopic metallicities determined by \citet{Layden1994}; for S10 we used photometric metallicities obtained from their Equation 23; and for V04 we used photometric metallicities computed as in M12. These photometric metallicity estimates are all consistent among themselves as they are based on the Fourier fitting technique of \citet{Jurcsik1996}, calibrated for \emph{ugriz} photometry for S10 and for Johnson-Cousins for V04 and M12. These are, in turn, guaranteed to be consistent with the \citet{Layden1994} spectroscopic metallicities, which were part of the calibration set used by \citet{Jurcsik1996} for the photometric metallicities.

\subsection{Survey completeness}\label{s:survey_completeness}

For RRLS surveys it is well known that the completeness and incidence of possible contaminants are difficult to estimate for \rrc~samples, which can be highly affected by period aliasing and are more prone to confusion with other types of short-period variable stars, such as W UMa eclipsing binaries and $\delta$ Scuti or SX Phe pulsating stars, specially at low galactic latitudes \citep[see e.g.][]{Kinman2010,Mateu2012}. We have therefore restricted our analysis to the 1305 RRLSs of type \typeab.

The knowledge of each survey's completeness is a key factor in the inference of density profile parameters.

The L95 sample suffers from distance-dependent and latitude-dependent incompleteness $I(R)$ and $I(b)$ respectively, which have been fully characterised by \citet{Layden1995} in their Appendix. We have included these incompleteness functions in our model as explained in Section~\ref{s:model_rhos}. 

\citet{Sesar2010} have estimated a uniform $100$ per cent completeness across their entire survey. For this sample, we have therefore assumed $I({\bf r})=1$ in our model from Section \ref{s:model_rhos}. We have also restricted the S10 sample to stars with $\alpha\geqslant345\degr$ and $|z|<20$ kpc, in order to avoid the known RRLS overdensities due to the Hercules-Aquila Cloud at $\alpha\sim330\degr$ and the trailing arm of the Sagittarius tidal stream at $z\sim -25$ kpc and $\alpha\sim30\degr$ \citep{Sesar2010,Ivezic2003,Belokurov2007b}. 

\citet{Vivas2004} and \citet{Mateu2012} have estimated the mean completeness of their surveys as a function of right ascension, as shown in their Figures 10 and 9 respectively. We have restricted the M12 survey to $\delta>-4.5\degr$, which ensures a uniform mean completeness of $I({\bf r})=0.98$ across the entire area covered by both surveys. We have also restricted the V04 sample to $\alpha<170\degr$ to avoid the RRLS excesses due to the Virgo Stellar Stream at $\alpha\sim 175\degr$ and the Sagittarius tidal stream at $\alpha\sim200\degr-240\degr$.

\subsection{Contamination by the thin disc and the bulge \label{s:contamination}}

As mentioned in Section~\ref{s:model_rhos}, our model does not include either a thin disc nor a bulge component.

The exclusion of a thin disc component is supported by the study by \citet{Martin1998} of a local sample of RRLSs with full space velocities and metallicities, in which the authors were unable to isolate a thin disc population. Given the small number of their sample (26 stars), \citet{Martin1998} do not discard completely that some stars from this component may exist. In any case, if there are indeed thin disc RRLSs, they would be only a very minor contributor compared with the number of halo and thick disc stars. Another, more recent piece of evidence is based on (photometric) metallicities of a large sample of RRLSs in the galactic plane from the VVV survey. In this work, \citet{Minniti2017b} also conclude that there are no metal rich RRLSs that can be associated with the thin disc of the Galaxy. However, in the future \emph{Gaia} should be able to provide strongest constraints on the existence, or not, of RRLSs in the thin disc.

We can also safely omit any bulge contribution. Although our sample contains several stars in the range $R<3$~kpc (see Figure~\ref{f:survey_coverage_aitoff}, right) none of them are expected to be associated with the bulge. The stars in this range of cylindrical radius $R$ come from the QUEST V04 sample, which has galactic latitude $30\degr\lesssim b \lesssim 50$ at $|l| \lesssim 15\degr$. Consequently, none of these stars are closer than $z=3.5$ kpc above the plane, which is $\sim 10$ times the  vertical scale-length of the bulge \citep{Zoccali2016}. We also discard the possibility of having stars belonging to the long bar associated with the bulge, which extends up to $\sim 5$ kpc from the Galactic center but it is well constrained to the plane \citep[180pc,][]{Wegg2015}. Even more, there is accumulating evidence that RRLSs do not trace the bar \citep{Dekany2013,Gran2016}.

\subsection{Survey Volume}\label{s:survey_volume}

The computation of the integral in Eq. \ref{e:nrr_model} requires a proper definition of the survey volume, with limits within which the completeness of the survey is known. This volume can be defined by the limits in $\alpha$ and $\delta$, as well as in heliocentric distance $R^o_{\hel}$ and $R^f_{\hel}$ which are obtained from each survey's saturation and completeness magnitude limits, plus the effect of the extinction. 

As illustrated by Figure 12 in \citet{Mateu2012},  the extinction is highly variable in the M12 survey due to the dust in the Galactic Plane. Since the saturation and completeness magnitude limits are fixed, this extinction variations will translate as varying distance limits as a function of the line of sight. To take this into account, the QUEST survey was divided into subregions chosen so as to keep extinction variations reasonably low ($\sigma_{A_V}\lsim0.2$ or $<10\%$ variation in distance). In each of these subregions the mean extinction was computed and, from the saturation and completeness magnitudes, the corresponding heliocentric distance limits derived. Table \ref{t:subregs} summarises the volume definitions adopted for each of the three RRLS surveys used.

Finally, we have also restricted the survey volume for all catalogues to within 25~kpc of the Galactic center, as mentioned in Sec.~\ref{s:halo_models}, to avoid modelling the contribution of the outer halo.

\subsubsection{Other RRLS catalogues}

Although many other RRLS catalogues are available in the literature, these were not used in our analysis mainly because of insufficient information about their selection function and formal overall incompleteness: The NSVS \citep{Kinemuchi2006} and ASAS \citep{Pojmanski2002} catalogues  have no estimate of completeness; the SEKBO \citep{Keller2008} and Catalina surveys \citep{Drake2013a,Torrealba2015} have relatively low completeness ($\gtrsim60\%$) and also do not report a formal incompleteness or selection function to account for variations across the survey; \citet{Miceli2008} gives a detailed account of the LONEOS-I survey completeness, but this could not be implemented in our analysis based on published data. Pan-STARRS is potentially an excellent source for this kind of studies but, in their latest catalogue of RRLSs, the completeness has not been characterized at low galactic latitudes \citep{Sesar2017}. Finally, the VVV survey \citep{Gran2016}, although with a large coverage at low latitude, was excluded as it is focused on the Galactic bulge and covers only the Galactic centre, out of the scope of the present study.

\begin{table}
\begin{center}
\begin{footnotesize}
\begin{tabular}{ccccccc}
\hline
Survey & $\alpha_o$ & $\alpha_f$ & $\delta_o$ & $\delta_f$  & $R^o_{\hel}$ & $R^f_{\hel}$  \\
                  & ($\degr$) & ($\degr$) & ($\degr$) & ($\degr$) & (kpc) & (kpc)  \\
\hline
QUEST & $60$ & $70$ &  $-4.5$   & $+0.1$ &  $3.6$ & $39.4$ \\
 & $ 70$ &  $ 83$ &  $-4.5$ & $+1.5$ & $3.3$ & $36.0$ \\
 & $ 73$ &  $ 83$ &  $-1.5$ & $+6.1$ & $3.4$ & $37.0$  \\
 & $ 83$ &  $ 93$ &  $+1.5$ & $+6.1$ & $1.7$ & $18.1$ \\
 & $ 90$ &  $100$ &  $-4.5$ & $-2.0$ & $2.0$ & $20.0$ \\
 & $100$ &  $113$ &  $-4.5$ & $-3.5$ & $1.4$ & $15.9$ \\
 & $113$ &  $120$ &  $-4.5$ & $-2.0$ & $3.3$ & $36.3$  \\
 & $120$ &  $135$ &  $-4.5$ & $+0.1$ & $3.6$ & $40.0$ \\
 & $135$ &  $150$ &  $-4.5$ & $+0.1$ & $3.6$ & $39.5$ \\
 & $193$ &  $210$ &  $-2.3$ & $+0.1$ & $3.6$ & $39.9$ \\
\hline
S10 &     $0$ & $60$  & $-1.2$ & $+1.2$ & $3.6$ & $40$ \\
       & $345$ & $360$ & $-1.2$ & $+1.2$ & $3.6$ & $40$ \\
\hline
L95 &     $0$ & $360$ & $-90$ & $+90$ & $0.5$ & $2.3$ \\
\hline
\end{tabular}\end{footnotesize}
\caption{Subregions defined for the computation of heliocentric distance limits in the QUEST, S10 and L95 catalogues. The L95 survey additionally has the restriction $|b|>10\degr$, to account for the avoidance zone set by \citet{Layden1995}.}
\label{t:subregs}
\end{center}
\end{table}

\section{Results for the halo and normal thick disc Model}\label{s:res_halo_normal_tkd}

\begin{figure*}
\begin{center}
 \includegraphics[width=2\columnwidth]{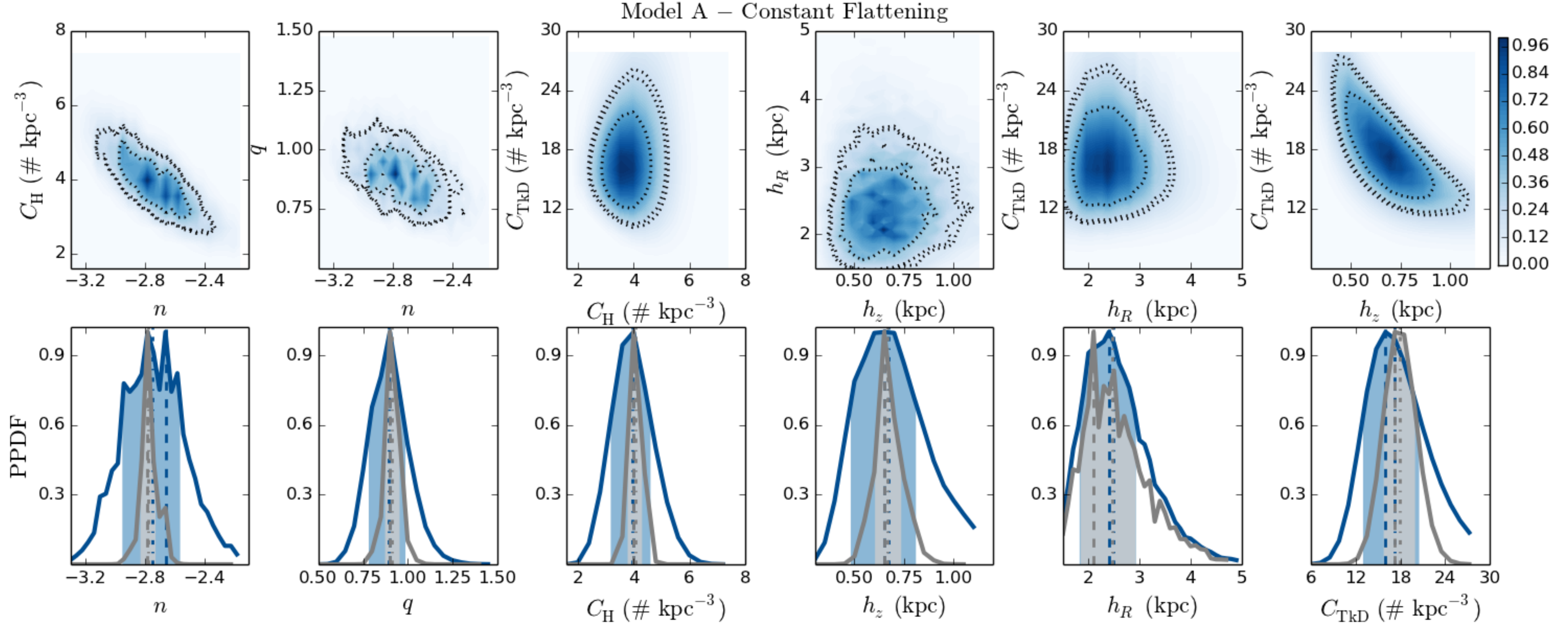}
 \caption{Posterior PDFs for the combined RRLS samples (L95+S10+QUEST), using the halo model with constant flattening and normal thick disc model. Top row: Marginal 2D posterior PDFs (the full set of 2D posterior PDFs is shown in Figure~\ref{a:sql_all_pdfs}). The dotted lines correspond to 1,2 and 3--$\sigma$ confidence regions. Bottom row: The dark (blue) and light grey lines show respectively the marginal 1D posterior and the joint 1D posterior, conditional on the best-fitting values for the remaining parameters. The dashed and dotted lines denote each distribution's mode and median respectively, and the shaded regions correspond to 1--$\sigma$ confidence intervals.}
\label{f:sql_qct_ppdfs}
\end{center}
\end{figure*}

\begin{figure*}
\begin{center}
 \includegraphics[width=2\columnwidth]{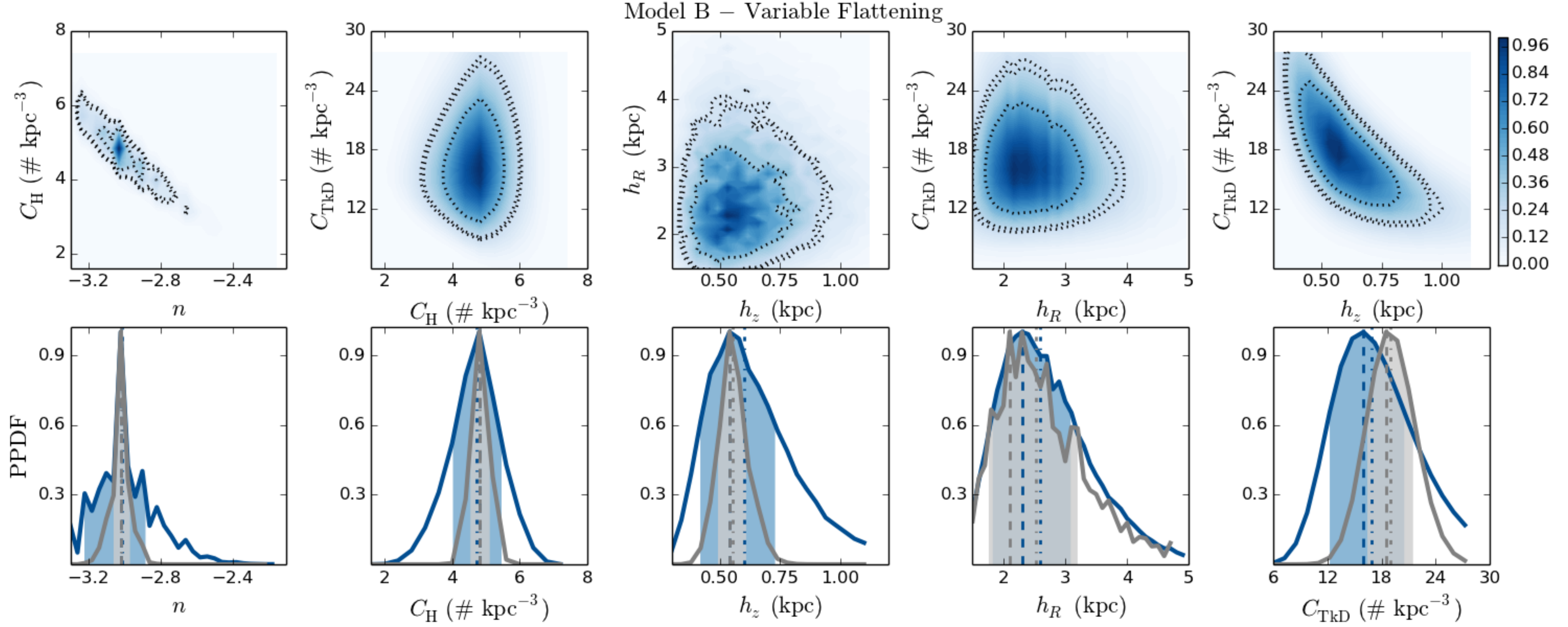}
 \caption{Posterior PDFs for the combined RRLS samples (L95+S10+QUEST), using halo Model B (variable P91 flattening) and normal thick disc model. Top row: Marginal 2D posterior PDFs (the full set of 2D posterior PDFs is shown in Figure~\ref{a:sql_all_pdfs}). The dotted lines correspond to 1,2 and 3--$\sigma$ confidence regions. Bottom row: The dark (blue) and grey lines show respectively the marginal 1D posterior and the joint 1D posterior, conditional on the best-fitting values for the remaining parameters. The dashed and dotted lines denote each distribution's mode and median respectively, and the shaded regions correspond to 1--$\sigma$ confidence intervals.}
\label{f:sql_qp_ppdfs}
\end{center}
\end{figure*}

\begin{table*}
\begin{center}
\caption{Best-fitting halo and normal thick disc parameters. The median, mode and $\pm1\sigma$ confidence intervals are reported for the marginal and joint posteriors.}\label{t:bestfits}
\begin{tabular}{ccccccccccc}
\hline\hline
& & \multicolumn{4}{c}{Marginal 1D PPDF} & \multicolumn{4}{c}{Joint PPDF} &\\
  \cmidrule(r{0.5em}l{0.5em}){3-6} \cmidrule(r{0.5em}l{0.5em}){7-10}
Model & Parameter&  Median & Mode &    $-1\sigma$ &  $+1\sigma$ &   Median & Mode & $-1\sigma$ &  $+1\sigma$  & Units\\ 
\hline
\multirow{6}{*}{Model A}  &        $q$ &   0.89 &   0.90 &  -0.12 &   0.09 &   0.91 &   0.90 &  -0.03 &   0.05 & \\
&        $n$ &  -2.75 &  -2.66 &  -0.29 &   0.10 &  -2.77 &  -2.78 &  -0.05 &   0.05 \\
 & $C_\h$  &   3.95 &   4.00 &  -0.82 &   0.61 &   4.06 &   4.00 &  -0.21 &   0.30 & \rrabkpc  \\
&    $h_z$  &   0.67 &   0.65 &  -0.17 &   0.16 &   0.67 &   0.65 &  -0.05 &   0.09  & kpc \\
&     $h_R$  &   2.49 &   2.40 &  -0.57 &   0.53 &   2.46 &   2.10 &  -0.25 &   0.82 & kpc \\
&$C_\tkd$  &   17.20 &  16.00 &  -3.02 &   4.52 &  17.92 &  17.25 &  -1.50 &   2.72 & \rrabkpc \\
\hline
\multirow{5}{*}{Model B}  &     $n$ &  -3.01 &  -3.02 &  -0.20 &   0.14 &  -3.02 &  -3.02 &  -0.04 &   0.05 & \\
& $C_\h$   &   4.71 &   4.80 &  -0.78 &   0.66 &   4.82 &   4.80 &  -0.26 &   0.31 & \rrabkpc \\
&    $h_z$  &   0.60 &   0.54 &  -0.12 &   0.19 &   0.55 &   0.54 &  -0.05 &   0.07 & kpc \\
&     $h_R$ &   2.59 &   2.30 &  -0.47 &   0.79 &   2.53 &   2.10 &  -0.34 &   1.10 & kpc \\
&$C_\tkd$ &  16.90 &  16.00 &  -3.74 &   4.52 &  19.03 &  18.50 &  -2.07 &   3.00 & \rrabkpc  \\
\hline
\end{tabular}
\end{center}
\end{table*}

The posterior PDFs of the combined L95+S10+QUEST RRLS sample, for the normal thick disc and halo Models A (constant $q$) and B (P91 variable flattening model), are shown respectively in Figures~\ref{f:sql_qct_ppdfs} and \ref{f:sql_qp_ppdfs}. In the figures, the top row shows 2D marginal posterior PDFs and the contours correspond to 1, 2 and 3-$\sigma$ confidence regions\footnote{These are taken as confidence intervals around the mode enclosing 68\%, 95\% and 99.5\% of the total probability, as a common analogy to the Gaussian case.}. The first two plots from the left show a combination of halo parameters; the third one shows the two normalizations for the halo and thick disc; and the last three plots, a combination of thick disc parameters (the full set of all 2D marginal posterior PDFs is shown in Figure~\ref{a:sql_all_pdfs}). The bottom row shows two types of 1D posterior PDFs: the marginal (blue) and the joint posteriors (light grey). For a given parameter, each of the 1D joint posteriors corresponds to the full (6D) posterior evaluated at the best-fitting value (mode) of the remaining five parameters. The dashed and dotted lines correspond respectively to the mode and median of each of the PDFs shown. The shaded regions correspond to the respective 1-$\sigma$ or 68\% confidence intervals, i.e. intervals around the mode containing 68\% of the probability, in 1-D for the marginal case and in 6-D for the joint posterior.  
Table~\ref{t:bestfits} (top) summarises the mode, median and 1-$\sigma$ confidence intervals for the marginal and joint posterior PDFs which are also shown in Figures~\ref{f:sql_qct_ppdfs} and \ref{f:sql_qp_ppdfs}. The maximum a posteriori values for halo Models A and B are respectively $lnP_A=-424.553$ and $lnP_B=-424.343$.

Figure~\ref{f:sql_qct_ppdfs} and \ref{f:sql_qp_ppdfs} illustrate the posteriors are very well behaved. For both models, most two-parameter posterior PDFs show fairly elliptic contours with little correlation between the parameters (see also Figure~\ref{a:sql_all_pdfs}). The notable (and expected) exceptions are the $p(n,C_\h)$ and $p(h_z,C\tkd)$ posteriors, which show strong anti-correlations that have been widely reported and discussed in the literature \citep{Robin1996,Robin2000,Chen2001}. These correlations between the scale ($n,h_z,h_R$) and normalization parameters ($C_\h,C_\tkd$), stem from the fact that $C_\h$ and $C_\tkd$ correspond to the \emph{local} densities of halo and thick disc RRLSs at $R=R_\odot$ and so, a profile with a steeper decline (i.e. smaller scale parameters) but larger normalization can produce the same number of RRLSs in a given volume as a profile with a slower decline and larger normalization. For the halo, 
there is only a very mild correlation between $n$ and the flattening $q$ and a slightly stronger one between $h_z$ and $n$, also expected given the halo flattening in the models is precisely in the z-direction. The two normalizations $C_\h$ and $C_\tkd$ are uncorrelated, with the halo normalization being much better constrained than the thick disc one. For the thick disc, the scale length and height are quite uncorrelated. The normalization and scale length posterior shows only very slightly skewed contours. The normalization and scale height posterior --in the right-most panel-- shows the mentioned anti-correlation with more banana-shaped contours.

The 1D posterior PDFs in these figures show \emph{for all parameters the modes of the marginal and joint PDFs are very similar}, particularly for Model A, and the overall shapes are very similar as well, aside from the expected fact that the joint PDFs are narrower. This is remarkable as it shows the full 6D posterior must be fairly elliptically-shaped, as opposed to very skewed or banana-shaped which would tend to make the joint PDF differ from the marginal PDF (as we will see in the next Section and Figure~\ref{f:sql_qp_ppdfs}). As seen in the Table and Figures, the mode and medians coincide quite well for the halo parameters and differ only very little for the three thick disc parameters, in all cases the difference being well within the corresponding joint-PDF 1-$\sigma$ confidence intervals. 

Figure~\ref{f:sql_qct_ppdfs} shows, in addition to overall agreement between the marginal and joint PDFs, the modes and medians also agree remarkably well, particularly for the halo parameters. For the thick disc, halo Model A  favours a relatively short scale height, with joint at marginal posterior modes coinciding at $0.65$ and medians coinciding at $0.67$~kpc respectively. The scale length, which has the most skewed distribution shows a slightly larger discrepancy between mode and median, although both favour also a short scale length: the joint and marginal posterior modes are $2.10$ and $2.40$ kpc,  and the medians $2.46$ and $2.49$ kpc, respectively.

\begin{figure}
\begin{center}
 \includegraphics[width=\columnwidth]{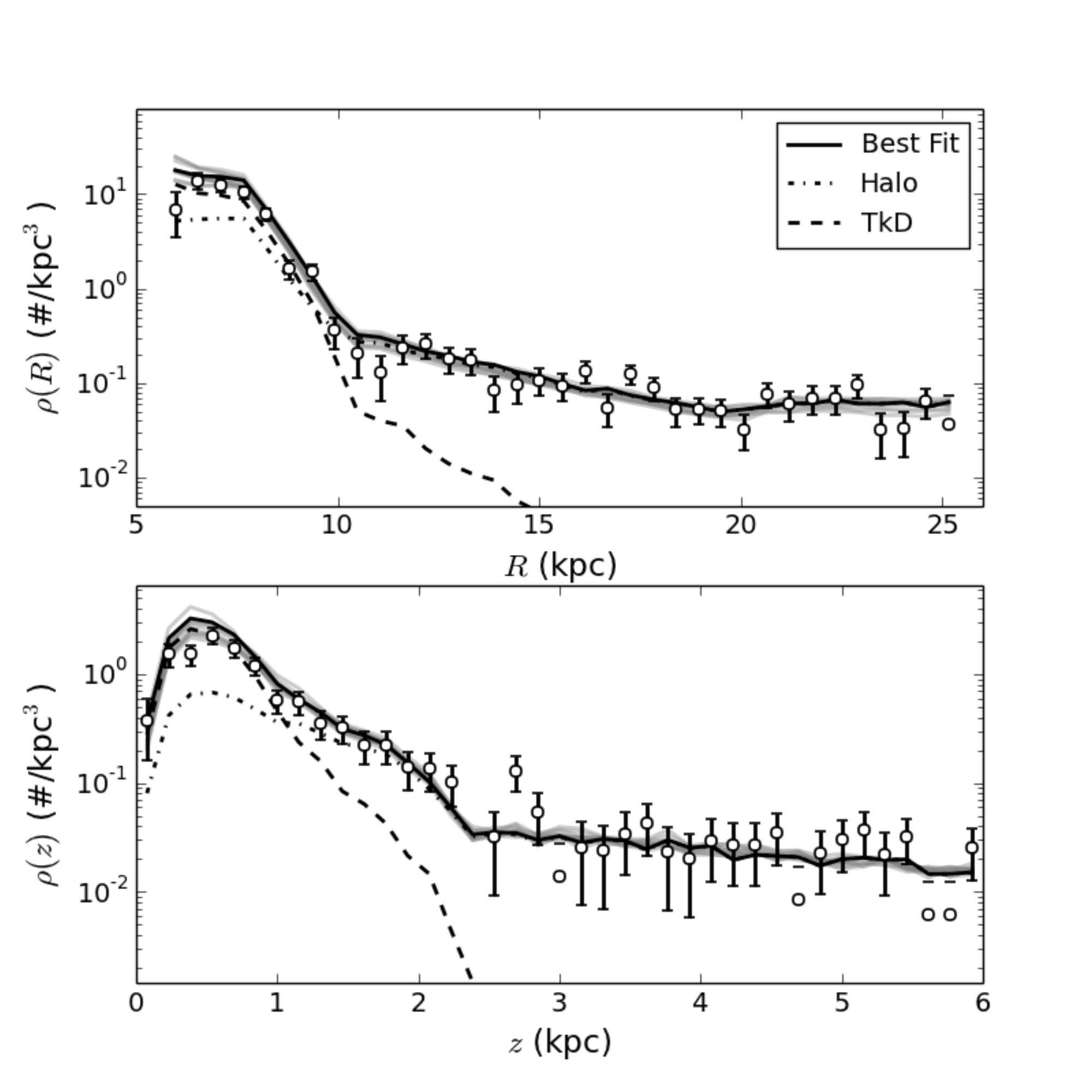}
 \caption{Average density as a function of $R$ (top) and $z$ (bottom), for the exponential disc model with halo Model A (constant flattening). The points represent the combined RRLS samples (L95+S10+QUEST), the solid and gray lines represent respectively the best-fitting model and 15 posterior samplings.}
\label{f:sql_qct_rhofits}
\end{center}
\end{figure}

\begin{figure}
\begin{center}
 \includegraphics[width=\columnwidth]{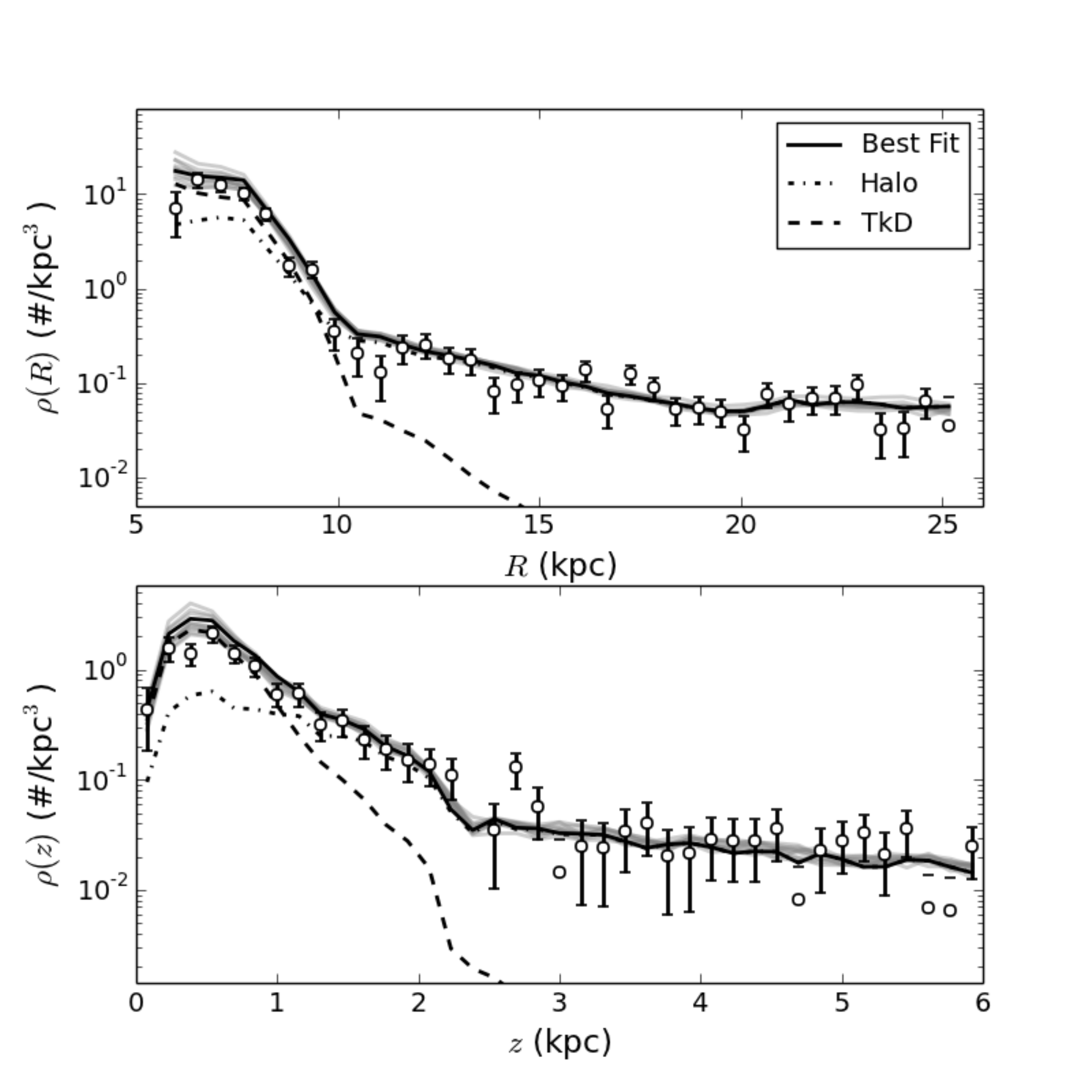}
 \caption{Average density as a function of $R$ (top) and $z$ (bottom), for the exponential disc model with halo Model B \citep[][variable flattening]{Preston1991}. The points represent the combined RRLS samples (L95+S10+QUEST), the solid and gray lines represent respectively the best-fitting model and 15 posterior samples.}
\label{f:sql_qp_rhofits}
\end{center}
\end{figure}

Figure~\ref{f:sql_qp_ppdfs} and Table~\ref{t:bestfits} show that for Model B all model parameters are also well constrained.
For the halo parameters the marginal posterior modes and medians for $n$ and $C_\h$ agree well with their joint posterior counterparts. For the thick disc parameters there is also good agreement, with only very slight discrepancies. 
For the scale height, halo Model B favours a slightly shorter value, with coinciding joint and marginal posterior modes of $h_z=0.54$ kpc, and medians only slightly larger (0.55 and 0.60 kpc). For the scale length, the joint and marginal posterior modes are respectively $h_R=2.1$ and $2.3$ kpc, while the medians ($2.53$ and $2.59$ kpc) show a smaller discrepancy with one another and favour slightly larger values. In both cases, these differences are well within the 1-$\sigma$ confidence intervals reported for $h_R$,  the parameter with the largest error bars. These results are remarkably consistent with those from halo Model A, hence, \emph{both halo models favour a relatively short scale length $\sim2.1$} kpc.  The (marginal) median thick disc normalisations obtained for Models A and B are also very similar, respectively $C_\tkd=16.90$ and $17.20$ \rrabkpc, and the modes coincide at $16.00$ \rrabkpc.
 
 To illustrate how the models agree with the data, Figure~\ref{f:sql_qct_rhofits} and \ref{f:sql_qp_rhofits} show, respectively for Models A and B, the \emph{average} density as a function of $R$ (top panel) and $z$ (bottom panel) for the combined L95+S10+QUEST sample (circles), the best-fitting (mode) model (black solid line) and 15 models drawn from the posterior PDF (grey transparent lines).  Additionally, for the best-fitting model, the halo and thick disc contributions are shown separately with the dotted and dashed lines respectively. Note that these average density profiles are computed \emph{including} the selection function, which accounts for features like the density decrease towards $z=0$ or the plateau for $R\lesssim7$. The figures show the models, including these features introduced by the selection function, fit the data very well overall; there are no strikingly obvious differences or improvements in one model fit with respect to the other model. This is also reflected in the similar values of the log-posterior probabilities, which translate into practically identical reduced $\chi^2$ values of $\chi_\nu^2=1.846$ and $\chi_\nu^2=1.845$. Given this, we find it unnecessary to go into a detailed model comparison, typically used to decide whether a model with more parameters and better goodness-of-fit results is significantly better than one with fewer parameters but slightly worse goodness-of-fit. Instead, our goal in using the two halo models is to emphasize that \emph{our main results for the thick disc parameters, both in the normal and flared disc models, are independent of the halo model used}. We will therefore, avoid making any claims as to which halo model is favoured by our data and leave this for a later analysis that may use a larger sample of RRLSs. 
 
\section{Results for the flared thick disc model}\label{s:res_flared_disk}

\begin{figure*}
\begin{center}
 \includegraphics[width=2\columnwidth]{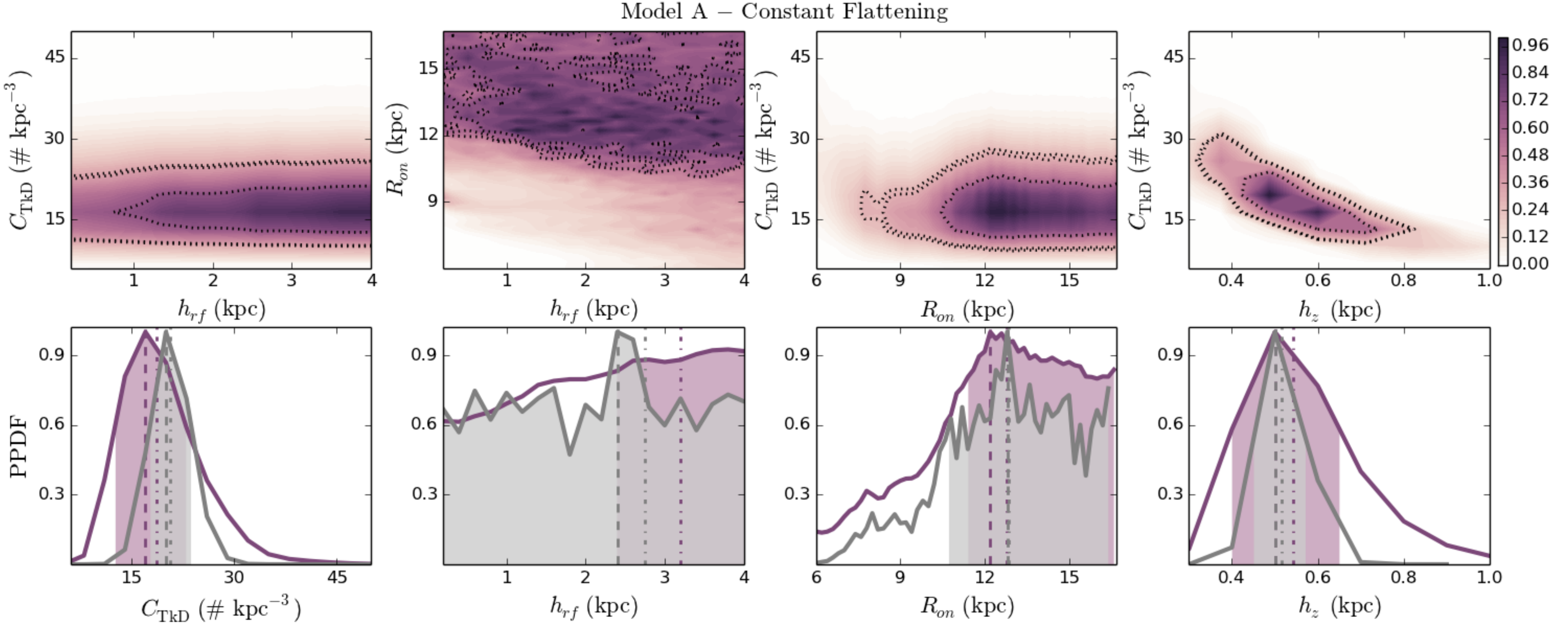}
 \includegraphics[width=2\columnwidth]{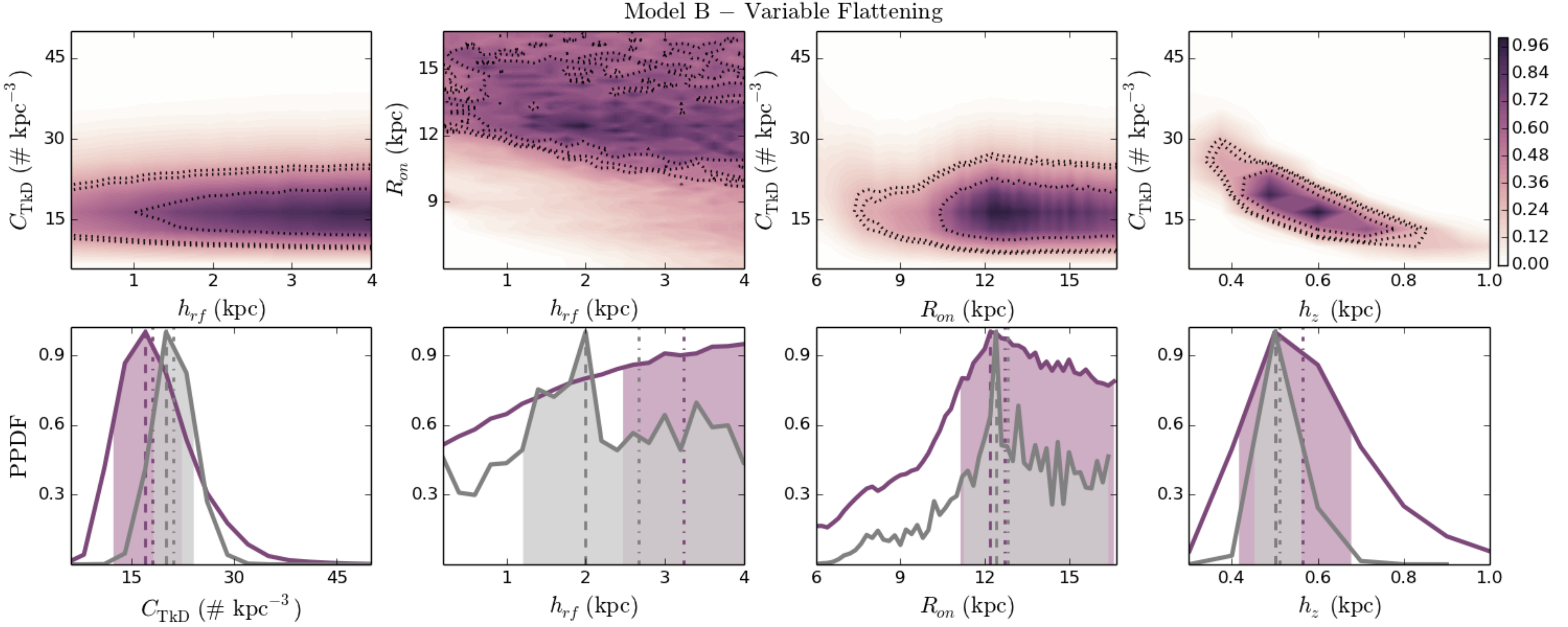}
 \caption{Posterior PDFs for the combined RRLS samples (L95+S10+QUEST) and the flared thick disc model, using best-fitting parameters for halo Models A (\emph{top panel}) and B (\emph{bottom panel}). Top row: Marginal 2D posterior PDFs. The dotted lines correspond to 1,2 and 3--$\sigma$ confidence regions. Bottom row: The dark (purple) and light grey lines show respectively the marginal 1D posterior and the joint 1D posterior, conditional on the best-fitting values for the remaining parameters. The dashed and dotted lines denote each distribution's mode and median respectively, and the shaded regions correspond to 1--$\sigma$ confidence intervals.}
 
\label{f:sql_flare_ppdfs}
\end{center}
\end{figure*}

The flared thick disc model posterior PDF was also computed by direct evaluation, as described in Section~\ref{s:nrr_computation}, assuming the best fit values found in the previous section for the halo parameters of Models A and B (Table~\ref{t:bestfits}). Figure~\ref{f:sql_flare_ppdfs} shows the 2D marginal posterior PDFs (1st and 3rd row) and the 1D marginal and joint posterior PDFs (2nd and 3rd row) for the flare model parameters, for halo Models A (\emph{top})  and B (\emph{bottom}). 

The best-fitting, as well as the marginal, values found for $C_\tkd$ and local scale height $h_z^\odot$, respectively $20$~\rrabkpc and $\sim0.6$~kpc, are very similar and consistent within the errors with their counterparts found for the normal thick disc model, with both halo Models. The marginal posteriors in Figure~\ref{f:sql_flare_ppdfs} show the flare model parameters $R_{on}$ and $h_{rf}$ are only mildly constrained. Both halo models favour an onset radius $R_{on}\sim12$~kpc with large uncertainty, particularly towards larger radii which cannot be ruled out at the 1-$\sigma$ confidence level. 

As in the case for the normal thick disc, the two halo models give very similar $\chi_\nu^2$ values of 1.020 and 1.026 respectively for models A and B, which again shows both models provide equally good fits of the data. This smaller $\chi_\nu^2=1.02$ values compared to the $\chi_\nu^2=1.8$ normal thick disc model, support the necessity of the flare parameters to model the thick disc profile. To make this comparison accounting for the difference in the number of model parameters, we compute the Bayesian Information Criterion $\mathrm{BIC}=\ln L -k\log{n}/2$  following \citet{Robin2014}, as defined by \citet{Schwarz1978}. For the flared disc using halo models A and B we get $\mathrm{BIC}=$-193.1 and -195.5 respectively, compared to the notoriously smaller values of $\mathrm{BIC}=$-424.3 and -424.6 for the normal disc. This confirms our first impression based on $\chi_\nu^2$ that \emph{the flared thick disc model accounts for the data significantly better than the normal disc model, and the different halo models make no appreciable difference}.
 
The joint posterior PDF on the flare parameters gives a best-fitting flare scale length $1.95$ kpc for halo B, with very large uncertainty, as shown by Table \ref{t:bestfits_flare} and illustrated by the 2D confidence regions of $p(h_{rf},R_{on})$ in Figure~\ref{f:sql_flare_ppdfs}. For halo Model A no significant mode is found for the joint posterior PDF, therefore it is not reported. Once the uncertainties in the remaining parameters are taken into account, the marginal 1D posterior for $h_{rf}$ shows only a lower limit of $\gtrsim1.8$ kpc can be reported for $h_{rf}$. 

The loose constraints found for the flare model parameters are reasonable results given that, as the right panel of Figure~\ref{f:survey_coverage_aitoff} shows, the coverage of the L95+S10+QUEST sample is limited to $R\lesssim15$ kpc at very low heights $|z|\lesssim3$~kpc. In any case, \emph{our results disfavour any flaring of the thick disc in the inner $\sim$11 kpc} and, if any, favour relatively mild flare models with scale length longer than $\gtrsim 1.8-2.4$ kpc and a flare onset radius $R_{on}\gtrsim 12$~kpc.  


\section{Discussion and comparison with previous studies}\label{s:comp_prev_aut}

\subsection{Halo}

\begin{figure}
\begin{center}
 \includegraphics[width=1.05\columnwidth]{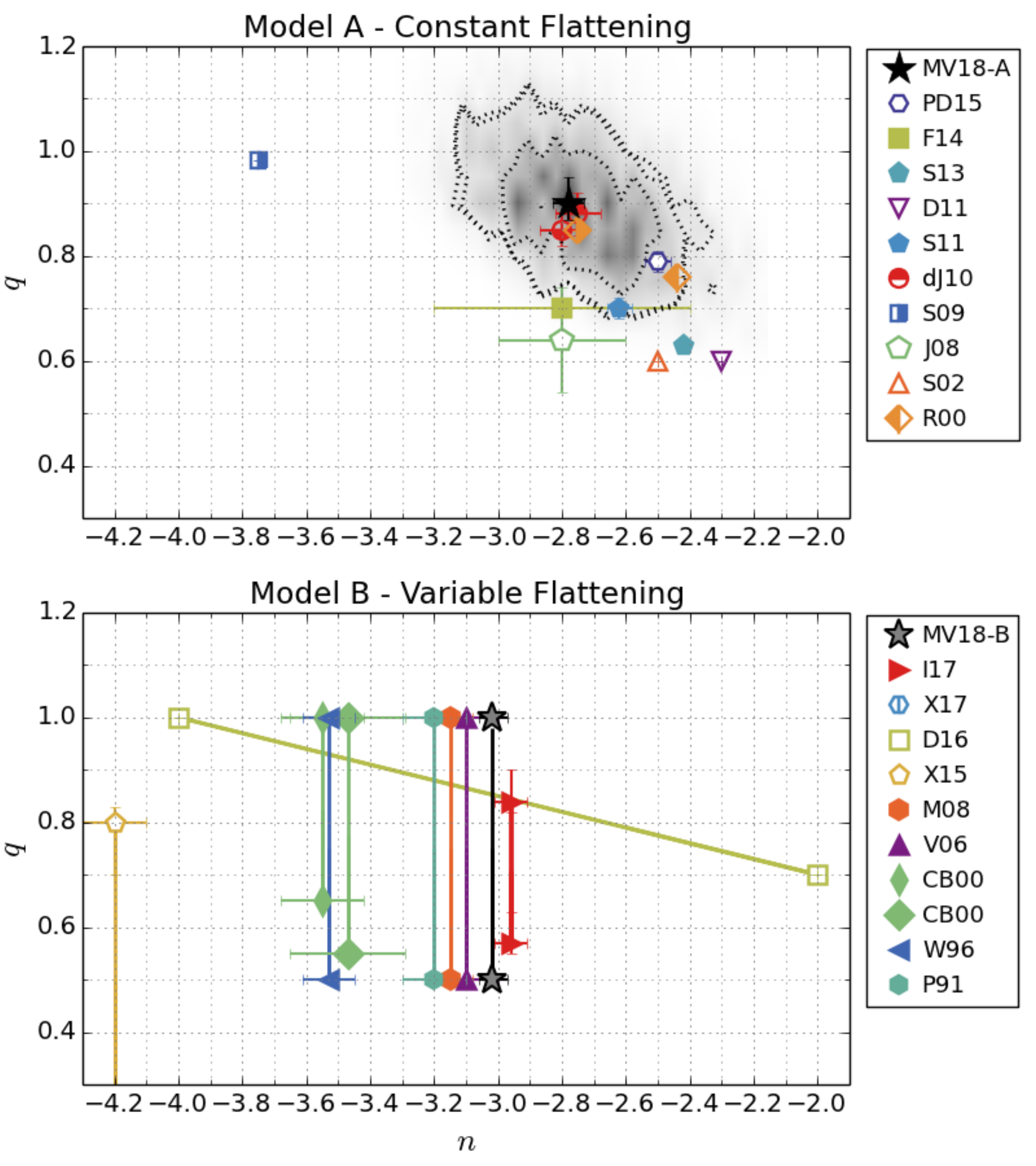}
 \caption{Comparison of results for the\emph{ inner halo} power law index $n$ and flattening $q$ in the literature, for a density profile model with constant flattening (\emph{top}) and variable flattening (\emph{bottom}). Results from previous works, summarised in Table~\ref{t:halo_lit}, are shown with different symbols: completely filled symbols correspond to results based on RRLSs, open symbols correspond to results from other tracers. Studies in which a particular value was assumed for $q$ are shown with high transparency. Our best-fitting values found for halo Models A and B are shown with the filled stars in the top and bottom panels respectively. In the top panel the marginal posterior $P(n,q)$ for halo Model A is shown in grayscale (same PDF as second panel top row of Fig.~\ref{f:sql_qct_ppdfs}) and the contours correspond to 1,2 and 3-$\sigma$ confidence regions. 
 In the bottom panel the lines connect ($n$,$q$) pair extremum values for each work, with $q$ always increasing as Galactocentric radius increases.}
 \label{f:n_q_prev}
\end{center}
\end{figure}

There have been numerous derivations of the halo structural parameters in the last decades, obtained using different tracers and techniques. Our results for the halo power law index $n$ and flattening $q$ are compared to those from previous studies in Figure~\ref{f:n_q_prev} and Table~\ref{t:halo_lit}. 

In Figure~\ref{f:n_q_prev} we compare: in the top panel, our result for halo Model A with those of studies that used a constant flattening $q$ as a free parameter; in the bottom panel, our result for halo Model B with those from previous studies that used a flattening law varying with radius. The different (coloured) symbols denote results from previous studies summarised in Table~\ref{t:halo_lit}, with filled symbols corresponding to results based on RRLSs. The top panel also shows the marginal posterior from Model A in grayscale (1,2 and 3-$\sigma$ confidence regions shown with the dotted lines), the best-fitting values for Models A and B are shown respectively in the top and bottom panels with the black and grey filled stars (referred to as MV18-A and MV18-B in the legend). In the bottom panel, the connected points mark the range of flattening values estimated for each study's model (shown in Table~\ref{t:halo_lit}), with flattening always increasing with radius.

For halo Model A (top panel) the plot shows only the works in which the authors have fit for $q$, summarised in the top part of Table~\ref{t:halo_lit}, to avoid giving disproportionate visual weight to studies that have set a fixed value for $q$, which are not directly comparable to ours. The main point this plot illustrates is \emph{the clear disagreement among all the different halo slope and flattening values found in the literature}, as there is no obvious concentration of points that favours  any particular $(n,q)$ pair. This disagreement is expected --to some extent-- because varying degrees of substructure are mapped along different lines of sight \citep{Bell2010} and thus expected to produce variations on the derived parameters of the density profile, as shown by \citet{Lowing2015}; this, coupled with likely different origins of the stellar populations producing the substructure, is expected to produce varying ratios in the number density of different tracers, as illustrated by \citet{Bell2010}. Excising known substructure like the Sagittarius and Virgo streams, as we have done in Section~\ref{s:survey_completeness}, is a common practice that  helps mitigate this, but variations are inevitably expected when comparing surveys with different area coverage and, especially, with different tracers. In the top panel of Figure~\ref{f:n_q_prev} then, the most straightforward comparison can be made with S11 and S13 (pentagons), the two results based on RRLSs with $q$ as a free parameter. The S11 result is consistent with ours at the 3-$\sigma$ level and, although the S13 result is outside our posterior PDF's 3-$\sigma$ contours, this result has no reported uncertainties, making it harder to asses how significant the discrepancy is. It is also worth noting that these two measurements seem to follow the general direction of the $n-q$ correlation of the posterior, which suggests the discrepancy can be due to the different spatial coverage, particularly in latitude, an effect \citet{Robin1996} has extensively illustrated.

For halo Model B (bottom panel), we can only compare results for the power law index, as $q$ has the fixed dependency with radius given by Eq.~\ref{e:qpreston}. Our best-fitting result $n=-2.98_{-0.08}^{+0.03}$ agrees within the 1-$\sigma$ error bars with results from P91, M08 and V06, all based on RRLSs and, particularly, with V06 based on high latitude QUEST RRLSs included in this work, but who use a different method to fit the density profile.  I17, in their recent work based on candidate RRLSs, use a more complex halo model and fit for the flattening as a function of radius. Their findings for $n=-2.96$ are remarkably close to our's, and in agreement within 1-$\sigma$ errors, in spite of the different dependence of flattening with radius between our models. CB00 and W96 results, based on blue horizontal branch stars and RRLSs respectively, favour a steeper power law $\sim-3.5$, discrepant with our results at $>$3-$\sigma$. 
Finally, the findings from D16, X15 and X17 (not shown), based on K giants, favour much steeper slopes $n<-4.$, which are entirely inconsistent with our results and with those of the remaining authors shown here, as well as with one another's. The fact that all but one (W96) of the significantly discrepant results for Model B are based on tracers other than RRLSs supports our previous argument that the different tracers are the main cause of the large spread of results from the literature discussed in Model A, aggravated by the fact that Model A has an extra free parameter that makes the spatial coverage play a more important role.

For the halo RRLS local normalization there are few results from previous authors. V06 reports $C_\h=6.6_{-0.7}^{+0.8}$~\rrabkpc for an assumed $q=0.6$ constant flattening (our halo Model A). This value is much higher than ours  $C_\h=4.0_{-0.21}^{+0.30}$~\rrabkpc, but given the discrepancy with our result for $q$ this disagreement is not surprising and is very likely due to correlation between these parameters. This is further supported by their result of $C_\h=4.2_{-0.4}^{+0.5}$~\rrabkpc for the P91 variable flattening law (our halo Model B), which agrees well with our best-fitting value within the error bars ($C_\h=4.8_{-0.26}^{+0.31}$\rrabkpc). For this same halo model, P91 found a normalization of 4.8~\rrkpc. This result includes RRLSs of all Bailey types, so accounting for the \rrab+\rrc~to \rrab~normalization factor $N_{ab+c}/N_{ab}=1.29$ (L95), this translates into a normalization of $3.7$~\rrabkpc which disagrees with our's and V06's results, although note the lack of reported uncertainties in the P91 result. It is interesting to note the strong dependency of $C_\h$ with $n$ and $q$, which translates into these results spanning values ranging over a factor of two, going from $3.6$ to $6.6$~\rrabkpc. 

\begin{table}
\begin{footnotesize}
\tabcolsep=0.1cm
\caption{Halo power law index and flattening obtained by different authors for the inner halo (Galactocentric radius $\lesssim$25~kpc). 
}\label{t:halo_lit}
\begin{tabular}{cccl}
\hline\hline
  \multicolumn{1}{c}{$n$} &
  \multicolumn{1}{c}{$q$} &
  \multicolumn{1}{c}{Tracer/Technique$^a$} &
  \multicolumn{1}{c}{Legend} \\
\hline
\multicolumn{4}{c}{Constant Flattening - Free $q$ } \\
\hline
\input{table_halo_authors.tex}
\hline
\end{tabular}
\raggedright \textsc{References:} {\textbf{MV18-A/B}=This Work - Model A/B};  PD15=\citet{PilaDiez2015}; F14 =\citet{Faccioli2014}; Z14=\citet{Zinn2014}; S13=\citet{Sesar2013}; D11=\citet{Deason2011}; S11=\citet{Sesar2011}; dJ10=\citet{deJong2010}; dJ10=\citet{deJong2010}; DP10=\citet{dePropris2010}; S09=\citet{Smith2009b}; J08=\citet{Juric2008}; V06=\citet{Vivas2006}; S02=\citet{Siegel2002}; R00=\citet{Robin2000}; R00=\citet{Robin2000};  C10=\citet{Carollo2010}; W09=\citet{Watkins2009}; M08=\citet{Miceli2008}; V06=\citet{Vivas2006}; I00=\citet{Ivezic2000}; W96=\citet{Wetterer1996}; S85=\citet{Saha1985}; X17=\citet{Xu2017}; I17=\citet{Iorio2017}; D16a,b=\citet{Das2016a,Das2016b}; X15=\citet{Xue2015}; M08=\citet{Miceli2008}; CB00=\citet{Chiba2000}; W96=\citet{Wetterer1996}; P91=\citet{Preston1991}\\
\raggedright $^a$ F-MSTOs= F type Main Sequence Turn-Off stars; BHB= Blue Horizontal Branch stars; MS= Main Sequence\\
\raggedright $^b$ Stars with $\FeH<-1.8$\\
\raggedright $^c$ Stars with $-1.6<\FeH<-1.1$\\
\end{footnotesize}
\end{table}

\subsection{Thick disc normalization}

In the literature there are few determinations of the local number density normalization for thick disc RRLSs. Based on a catalogue of GCVS4 RRLSs, L95 finds a mid-plane density normalization of $10\pm4$~\rrkpc, ~which, correcting by the same factor as before (previous section), translates into a normalization of $7.8\pm 3.1$~\rrabkpc~for type \typeab~RRLSs.  

On a more recent study, \citet{Amrose2001} measured the \emph{total} mean density in the solar neighbourhood, finding a value of $5.8\pm0.7$~\rrabkpc, based on a sample of \rrab~stars from the ROTSE-I survey. However, these results are not directly comparable to ours as they report the mean value for the \emph{total} density in the solar neighbourhood, whereas the density normalization represents a local value at exactly $R=R_\odot$ (as mentioned in Section~\ref{s:res_halo_normal_tkd}) which, depending on the particular shape and parameters of the density profile, will produce different values of the total number of RRLSs in a given volume, and therefore different values of the  average density. 

Our results for the thick disc normalization, $18.50$ and $17.25$~\rrabkpc~for halo Models A and B,  are much higher than the L95 results and incompatible at more than the $3\sigma$ level in both cases.  The scale height $h_z=0.7^{+0.5}_{-0.3}$ kpc found by L95 has large reported uncertainty and agrees within the errors with our results $h_z=0.65_{-0.05}^{+0.09}$ and $h_z=0.54_{-0.05}^{+0.07}$ kpc for halo models A and B, particularly the former. This rough agreement makes it unlikely that the $h_z-C_\tkd$ correlation can account for the discrepancy in $C_\tkd$. This discrepancy might stem, however, from the different the radial dependence of the density profile used in L95, since they do not fit for the disc scale length and assume an exponential shape for the halo, as opposed to the more common power-law assumed here. 

\subsection{Thick disc scale height and length}\label{s:res_hzhr}

\begin{figure}
\begin{center}
 \includegraphics[width=\columnwidth]{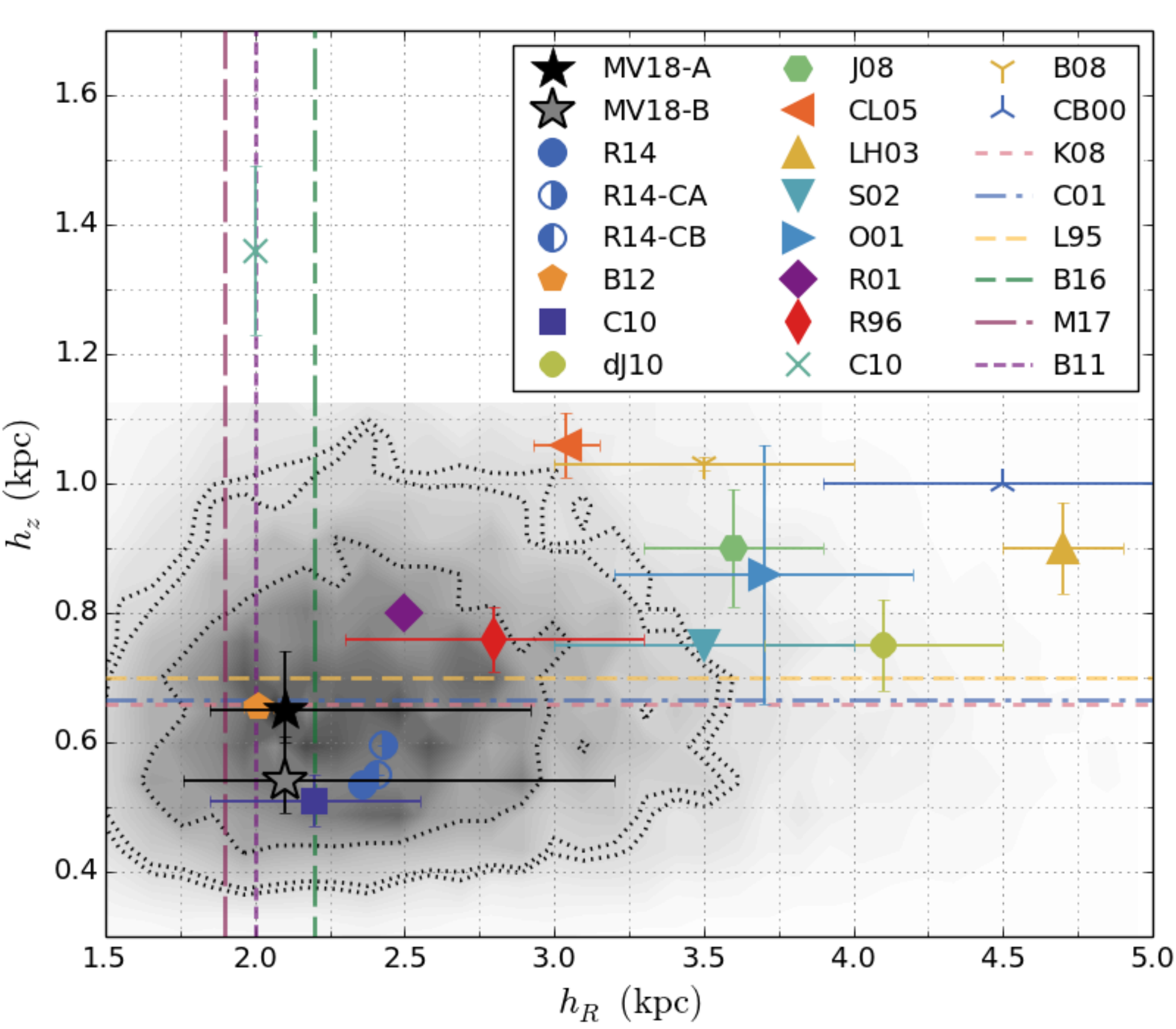}
 \caption{Comparison of results for the thick disc scale height $h_z$ and scale length $h_R$ in the literature. Results from previous works, summarised in Table~\ref{t:tkd_lit}, are shown with different (coloured) symbols. Open symbols correspond to estimates for the Metal-Weak thick disc. The vertical and horizontal lines represent results from works that report only  either $h_R$ or $h_z$ respectively. Note that results from R14, R14-CA/B and R01 lack reported error bars. Our best-fitting values found for the normal thick disc plus halo Models A and B are shown with the filled and empty (black) star, respectively. The marginal posterior $P(h_z,h_R)$ for halo Model A is shown in grayscale and the contours correspond to 1,2 and 3-$\sigma$ confidence intervals (same as fourth panel top row of Fig.~\ref{f:sql_qct_ppdfs}). }
\label{f:hz_hr_prev}
\end{center}
\end{figure}

In Figure~\ref{f:hz_hr_prev} we compare our results for the thick disc scale height and scale length with results from previous authors. The plot shows $h_z$ as a function of $h_R$ for the works summarised in Table~\ref{t:tkd_lit}, each shown with a different symbol. The three unfilled symbols correspond to estimates for the Metal-Weak thick disc \citep[see e.g.][]{Majewski1993}. The vertical and horizontal lines represent results from works that report only $h_R$ or $h_z$ respectively. Our best-fitting values found for the normal thick disc plus halo Models A and B are shown with the filled and empty (black) star, respectively. The marginal posterior $P(h_z,h_R)$ for halo Model A is shown in grayscale and the contours correspond to 1,2 and 3-$\sigma$ confidence intervals.

 Figure~\ref{f:hz_hr_prev} shows the distribution of results from different authors in the $h_z-h_R$ plane is fairly bimodal, clustering around relatively long scale lengths $\sim3.5$--$3.7$~kpc with long scale heights $\sim0.9$~kpc or shorter scale lengths $\sim2.3$ kpc with also shorter scale heights $\sim0.6$~kpc. \emph{Our results, independently of the halo Model, are compatible with the latter, a shorter scale length $\sim2.1$~kpc and scale height $0.6$~kpc}. The works from C10, B12, R14, R14-CA/B, R01 and R96 which have found different combinations of these short $h_z$-$h_R$, all lie within the $1\sigma$ confidence interval of our marginal posterior for Model A shown in the figure.  All of the works that predict longer $h_z-h_R$, namely S02, CL05, J08, dJ10, LH03, O01, B08 and CB00,  are inconsistent with our results at the $3\sigma$ level at least. It is interesting to point out, also, that \emph{our marginal posterior PDF shows no signs of a second peak at high $h_z-h_R$}. Note that, since our prior goes up to $h_R=5$~kpc (Table~\ref{t:priorlims}) and we have computed our posterior PDFs from direct evaluation, \emph{the uni-modality of the posterior is a strong result}.  Also, there is remarkable agreement in $h_R$ with B16 and B11 (vertical long and short dashed lines), and in $h_z$ with C01, L95 and K08 (horizontal dashed and dashed-dotted lines), the latter two based also on RRLS analyses. Note also, as shown in Table~\ref{t:tkd_lit}, that \emph{ours is the first determination of the thick disc scale length made with RRLSs.}
 
\begin{table}
\tabcolsep=0.14cm
\begin{footnotesize}
\caption{Thick disc scale lengths and heights obtained by different authors.}\label{t:tkd_lit}
\begin{tabular}{cccl}
\hline\hline
  \multicolumn{1}{c}{$h_R$} &
  \multicolumn{1}{c}{$h_z$} &
  \multicolumn{1}{c}{Tracer/Technique} &
  \multicolumn{1}{l}{Legend} \\
\hline
\input{table_tkd_authors.tex}
\hline
\end{tabular}
\raggedright \textsc{References:} {\textbf{MV18-A/B}=This Work - Model A/B}; M17=\citet{Mackereth2017}; B16=\citet{Bovy2016}; R14=\citet{Robin2014}; R14-CA=R14-\citet{Czekaj2014}-A; R14-CB=R14-\citet{Czekaj2014}-B; B12=\citet{Bovy2012}; B11=\citet{Bensby2011}; C10=\citet{Carollo2010}; dJ10=\citet{deJong2010}; J08=\citet{Juric2008}; CL05=\citet{CabreraLavers2005}; LH03=\citet{Larsen2003}; S02=\citet{Siegel2002}; C01=\citet{Chen2001}; O01=\citet{Ojha2001}; R01=\citet{Reyle2001}; R96=\citet{Robin1996}; K06=\citet{Kinemuchi2006}; L95=\citet{Layden1995}; C10=\citet{Carollo2010}; B08=\citet{Brown2008}; CB00=\citet{Chiba2000} \\
\raggedright $^a$ BHB= Blue Horizontal Branch stars; MS=Main Sequence, RC=Red Clump giants; RGB= Red Giant Branch stars
\end{footnotesize}
\end{table}

Our results are consistent with those of B12, B16, R14, R14-CA/B, R01, R96 and  C10, within the marginal uncertainties. These authors follow three different methodologies: C10 uses kinematic data and Jeans's theorem; B12, B16 and M17 use a Bayesian forward model to fit the density profile parameters for MAPs selected in the $\alphaFe$--$\FeH$ plane; R14, R01 and R96 follow a common population synthesis methodology, comparing star counts in observed and modelled Hess diagrams.

C10 uses a local sample ($R_{hel}<4$ kpc) of SEGUE  SDSS-II calibration stars with spectroscopic parallaxes, proper motions and radial velocities to infer the thick disc scale length and height from a Jeans theorem analysis. 
Their scale length of $h_R=2.2\pm0.35$, shown in Figure~\ref{f:hz_hr_prev} with a square, is well in agreement with ours within the uncertainties, with the scale height also coinciding remarkably with our results for halo Model B. Note, however, that C10 separates the thick disc into the canonical thick disc and the Metal-Weak thick disc components. For the latter they estimate a much larger scale height of 1.36~kpc which is at odds with the rest of the literature, even for the Metal Weak thick disc (see B08, CB00), yet it they also estimate it to have a short scale length of $\sim2$ kpc. 

B12, B16 and M17 use G-type SEGUE dwarfs and APOGEE Red Clump and RGB stars respectively to fit single exponential density profiles of MAPs, defined as bins in the $\alphaFe$--$\FeH$ plane. Their fitting methodology is the same as that of \citet{Lombardi2013} used here, extended to include a dependence of the selection function on the colour and metallicity of the stars. In their analysis of the spatial and kinematical structure and surface density of the MAPs, B12 and B16 argue against a distinct thin and thick decomposition of the Galactic disc.  
Using RRLSs as tracers we cannot argue about the adequacy of a distinct thin/thick disc decomposition, since RRLSs with thin disc kinematics are very scarce at best \citep{Martin1998}. However, our results can be compared with the \citeauthor{Bovy2012b} high-$\alphaFe$ low-$\FeH$ populations, i.e. the old and metal poor populations traced by RRLSs. For the high-$\alphaFe$ MAPs B12 and B16 find respectively $h_R=2.01\pm0.05$~kpc and $h_R=2.2\pm0.2$~kpc, which agree very well  with our estimate of $h_R=2.1_{-0.25}^{+0.82}$~kpc and $h_R=2.1_{-0.34}^{+1.10}$~kpc for halo Models A and B respectively. For their high-$\alphaFe$ sample, M17 finds a slightly smaller scale length $h_R=1.9\pm0.1$~kpc, still consistent with our result, although this corresponds to the mean for their full sample, which is relatively metal-rich ($\FeH>-0.6$). For the scale height, the B12 estimate $h_z=0.66\pm 0.013$~kpc is in perfect agreement with our result  from halo Model A $h_z=0.65_{-0.05}^{+0.09}$.  Although RRLSs are much more precise distance tracers than Red Clump stars and G-dwarfs, our uncertainties are higher than those of B16 and B12 due to the relatively low number of RRLSs in our analysis, particularly at simultaneously large radii and low latitude, as noted similarly by B16 when comparing their Red Clump results with the B12 G-dwarf results. As discussed before, large scale surveys that will increase the low latitude coverage of RRLSs will very soon greatly improve these uncertainties.

R14, R01 and R96 use the Besan\c{c}on Galactic Model to compare star counts in Hess diagrams for different photometric surveys and fit for the parameters of the thin and thick disc density profiles. In their most recent work, \citet{Robin2014} find a two-episode model for the thick disc provides the best fit in their stellar population synthesis study. The ages for their two formation episodes are 10-11 and 11-12 Gyr, and find a longer scale height and scale length for the older component compared to the younger one, both with a metallicity $z=0.003$ ($\FeH\sim-0.8$). \citeauthor{Robin2014} also note the younger component dominates the density parameters when the thick disc is modelled as a single-age stellar population. Since RRLSs exist only in stellar populations that are both old ($>10$ Gyr) and metal-poor ($\FeH \lesssim -0.5$), these stars trace their two formation episodes as a whole. So our results are best compared with those of \citet{Robin2014} for their main (`young') thick disc component, i.e. the dominant component, cited in Table~\ref{t:tkd_lit}. Our scale length of $h_R=2.1$~kpc  agrees well with the R14, R14-CA and R14-CB results  $h_R=2.36, 2.41$  and $2.43$ kpc respectively, while our scale height estimate $h_z=0.65$~kpc for Model A is slightly larger than their reported values $h_z=0.535\pm0.0046,0.596$ and $0.549$~kpc respectively, very similar to our result for Model B $h_z=0.54$~kpc, although note their latter two results are reported without uncertainties.

B12 claims the cause for the discrepancy between the short versus long $h_z-h_R$ results is precisely that geometrical decomposition analyses predict longer scale lengths because the component with the largest scale height gets paired with the largest scale length. \citet{Martig2016b} attribute this to the combined effect of a radial age gradient in the disc \citep[proposed by][]{Minchev2015}, which they observe with APOGEE giant stars with measured ages, and the flaring of the (predominantly young) outer disc.  They find the old component, which is also indeed $\alpha$-rich, is more centrally concentrated and dominates the inner disc, compared to the young and $\alpha$-poor component. 

Our results support this interpretation since our sample is a clear tracer of old (metal-poor) populations, as argued in Section~\ref{s:intro}. Furthermore, from our comparison with the literature we can see that our main result, \emph{that the RRLSs thick disc has short $h_z$-$h_R$}, agrees with \emph{all previous studies that do a chemical, as opposed to a geometrical,  decomposition}: i)  K08 and L95, the only two previous studies of the thick disc structure traced with RRLSs, which report only $h_z$; ii) the MAP analyses from B12, B16 and M17; iii) B11 who, based on high-resolution abundances for K giants, find a short scale length (2~kpc)  when decomposing their sample by $\alphaFe$-abundances, and iv) C10, who do a kinematic analysis of stars in different metallicity subpopulations. Out of the typically geometrical methods,  the R14 results stand out as the only ones that --in agreement with our's-- also find a shorter $h_z-h_R$, compared to previous results from the same group (R96,R01) with the same CMD fitting method, the main difference being that the thick disc in R14 is modelled with two formation episodes of slightly different ages.  All this further supports the conclusion that  a geometric decomposition of the disc tends to favour spuriously long $h_z-h_R$ combinations; while age and $\alpha$ abundance or MAP decompositions, based on properties a priori independent on the structural parameters to be inferred, provide a more consistent and physically motivated decomposition that favour a short $h_z-h_R$ pairing. 
   
\subsection{Thick disc flare}

Our results for the thick disc flare are in very good agreement with those of \citet{Robin2014} who report \emph{no evidence of flaring up to $\sim$11 kpc in the main thick disc formation episode (age $\sim$10-11 Gyr)}.  
Although they do find a significant flaring of their old (11-12 Gyr) thick disc component, as discussed in Sec.~\ref{s:res_hzhr}, our RRLS findings must be compared to their results for the thick disc as a whole, which is dominated by the `younger' (10-11 Gyr) component. B16 also find their high-$\alphaFe$ MAPs do not flare in the  radial range $4<R({\rm kpc})<14$ covered by their Red Clump sample. In approximately the same radial range,  \citet{Mackereth2017} used APOGEE RGB stars to analyse the flaring of MAPs as a function of  \emph{age}, in addition to metallicity. Their stars have $\FeH$ in the range $[-0.6,+0.2]$, so our RRLS results can only be compared to stars in their most metal-poor bin $\FeH <-0.5$ with ages $\gtrsim$10 Gyr. Their Figure~7 shows 
for the most metal-poor stars ($\FeH <-0.5$) the high-$\alphaFe$ subset in the age range $9<$age$<11$ shows no flaring  and only a very slight one in the next age range $11<$age$<12$, which is also consistent with our findings. 

\citet{Hammersley2011} obtained a flare onset radius $\Ron=16$ kpc with a flare scale length $\hrf=4.5\pm1.5$ kpc, also in good agreement with our results which favour $\Ron\gtrsim 11$ kpc and a large scale length $\hrf>3$ kpc.
\citet{Conn2012} find a shorter flare onset radius $\Ron=12.6$ kpc and also a shorter scale length $\hrf=2.1$ kpc. Their flare onset radius coincides remarkably well with our best-fitting marginal modes $\Ron=12.9$ and $12.4$~kpc found with halo models A and B although, as noted before, we cannot confidently rule out larger values for the onset radius. Our analysis does not allow drawing any firm conclusions about the flare scale length either, just to say that this value is not ruled out by our findings.

\citet{Momany2006} and \citet{Yusifov2004} studied the disc flaring using Red Giant Branch stars and pulsars.
Both tracers are found in populations ranging from very young to very old, and so, trace a combination of the thin+thick discs. Although the two tracers follow both discs, \citet{Yusifov2004} finds that pulsars trace a flare that follows more closely the shape of the dust and HI flare than that of the red giant flare from \citet{LopezCorredoira2002}, hence these results are less likely to reflect the characteristics of the thick disc flare. \citet{Momany2006} also find no flaring up to 15 kpc with a scale height $h_z\sim0.65$ kpc, both in very good agreement with our findings. Beyond that radius they find a mild flaring that they can trace out to 23 kpc.

\begin{table*}
\begin{center}
\caption{Best-fitting flared thick disc parameters. The median, mode and $\pm1\sigma$ confidence intervals are reported for the marginal and joint posteriors.}\label{t:bestfits_flare}
\begin{tabular}{ccccccccccc}
\hline\hline
& & \multicolumn{4}{c}{Marginal 1D PPDF} & \multicolumn{4}{c}{Joint PPDF} &\\
  \cmidrule(r{0.5em}l{0.5em}){3-6} \cmidrule(r{0.5em}l{0.5em}){7-10}
Model & Parameter&  Median & Mode &    $-1\sigma$ &  $+1\sigma$ &   Median & Mode & $-1\sigma$ &  $+1\sigma$ & Units\\ 
\hline
\multirow{4}{*}{Model A}  &   
$C_\tkd$          &  19.99 &  20.00 & -5.02 &   3.87 &    20.81 &  20.00 & -1.55 &   3.26 & \rrabkpc\\ 
 &  $\hrf$     &     2.33 &   3.50 & -1.73 &   0.30 &     2.49 &   $\cdots$ & -2.53 &   0.00 & kpc\\ 
 & $\Ron$    &  13.15 &  12.30 & -1.11 &   5.10 &    12.32 &  12.30 & -3.65 &   2.75 & kpc\\ 
  & $h_z$     &   0.66 &   0.65 & -0.13 &   0.12 &     0.65 &   0.65 & -0.04 &   0.03 & kpc\\ 
\hline
\multirow{4}{*}{Model B}  &   
$C_\tkd$ &  21.21 &  20.00 &  -5.22 &   6.02 &  22.54 &  23.00 &  -3.63 &   2.40 & \rrabkpc\\ 
  &  $\hrf$ &   3.18 &   5.20 &  -2.76 &   0.25 &   2.66 &   1.95 &  -1.06 &   3.50 & kpc \\ 
   & $\Ron$ &  12.39 &  12.40 &  -1.51 &   2.96 &  12.67 &  12.40 &  -1.29 &   2.30 & kpc \\ 
   & $h_z$ &   0.52 &   0.50 &  -0.11 &   0.12 &   0.50 &   0.50 &  -0.05 &   0.05 & kpc\\     
\hline
\end{tabular}
\end{center}
\end{table*}

\section{Conclusions}\label{s:conclusions}

In this work we have used three public RRLS~surveys to analyse the structure of the Galactic thick disc: the L95 compilation of nearby GCVS RRLSs; the deep high-latitude S10 RRLS survey in SDSS Stripe 82; and the low and high latitude QUEST RRLS surveys from \citet{Mateu2012} and \citet{Vivas2004}. A homogeneous catalogue was produced combining these, with all distances recomputed to be on the same scale. A Bayesian Poisson forward model was used to fit the parameters of the combined halo+thick disc density profile using two halo Models: A, with a constant flattening; and B, following the P91 variable flattening model. A flared thick disc model was also explored, assuming the best-fitting values found for the two halo Models. We summarise our results as follows:

\begin{itemize}

\item[--] The best-fitting (joint posterior mode) parameters found for halo Model A (constant flattening) are $n=-2.78_{-0.05}^{+0.05}$, $q=0.90_{-0.03}^{+0.05}$ with a normalization of $C_\h=4.00_{-0.21}^{+0.30}$ \rrabkpc. For halo Model B (P91 flattening law) the best-fitting parameters found are $n=-3.02_{-0.04}^{+0.05}$ with a normalization of $C_\h=4.8_{-0.26}^{+0.31}$ \rrabkpc.

\item[--] Our results for halo Model A are consistent with those of S11 and midly discrepant with S13, the only two studies made with RRLSs, in the inner halo and with a free $q$ parameter in this halo model, and so directly comparable to ours. For halo Model B, our results are consistent with all but one (W96) of the RRLS results from the literature (P91,M08,V06 and I17).

\item[--] For the normal thick disc model our results favour a short-$h_z-h_R$ combination, independently of the halo Model used. For halo Model A we find $h_z=0.65_{-0.05}^{+0.09}$~kpc, $h_R=2.1_{-0.25}^{+0.82}$~kpc (joint posterior mode). These are the first determinations of the thick disc scale length made to date using RRLSs.

\item[--] Our thick disc posterior PDFs are unimodal. These were computed by direct evaluation and, so, confidently rule out the existence of a second peak at large $h_z-h_R$, as suggested by some previous studies (S02, CL05, J08, dJ10, LH03, O01, B08 and CB00).

\item[--] Our findings for the thick disc normalization $C_\tkd=17.25_{-1.50}^{+2.72}$~\rrabkpc~for halo Model A (and $18.50_{-2.07}^{+3.0}$ for Model B), are much higher than the only previous result reported in the literature by L95, who found a value of $C_\tkd=7.8\pm3.1$ \rrabkpc.

\item[--] Although the flared thick disc model parameters could only be mildly constrained given the limited coverage of the RRLS samples in radius at low $z$, our results \emph{highly disfavour any flaring in the inner $\sim11$~kpc}.  Both halo models give a best-fitting value for the onset radius $\Ron\sim12.5$~kpc, but larger values of $\Ron$ cannot be ruled out. For the flare scale length only a lower limit $\hrf\gtrsim1.8$~kpc can be reported for halo Model B. 

\item[--] The flared disc model is found to provide a significantly better account of the data compared to the normal disc and independently of the halo model used, which has no appreciable effect on the goodness of fit.

\item[--] The short scale height and large normalization found for the normal thick disc model are also supported by the flared disc model, for which best-fitting values of $0.65$~kpc (at the solar radius) and 20 \rrabkpc~were found.

\end{itemize}

Our results provide an independent confirmation of findings from C10, B12, R14, R01 and R96 and from \citet{Bovy2012} for the metal-poor high-$\alphaFe$~disc, which favour a short scale height ($\sim0.6$~kpc) and short scale length ($\sim2.2$~kpc) for the thick disc. This is a robust confirmation of these results, as they are based in \rrab~star surveys known to be complete, uncontaminated, with negligible to no contribution from the thin disc and with precise individual distances. The methodology used here in combination with upcoming large RR Lyrae star samples from \emph{Gaia} and LSST will be a powerful tool to probe the outskirts of the galactic thick disc and give stronger constraints on the existence of a thick disc flare.

\section*{Acknowledgments}

We thank the anonymous referee for interesting comments and suggestions that resulted in a better manuscript. CM wishes to thank Nicolas Martin, David Hogg and Tom Loredo for useful suggestions and discussions, and Bolivia Cuevas for her help with bibliographic resources. CM acknowledges the use of TOPCAT \citep{Taylor2005} through out the course of this investigation.


\newpage
\appendix
\section{Marginal 2D posterior PDFs}

\begin{figure*}
\begin{center}
 \includegraphics[width=1.4\columnwidth]{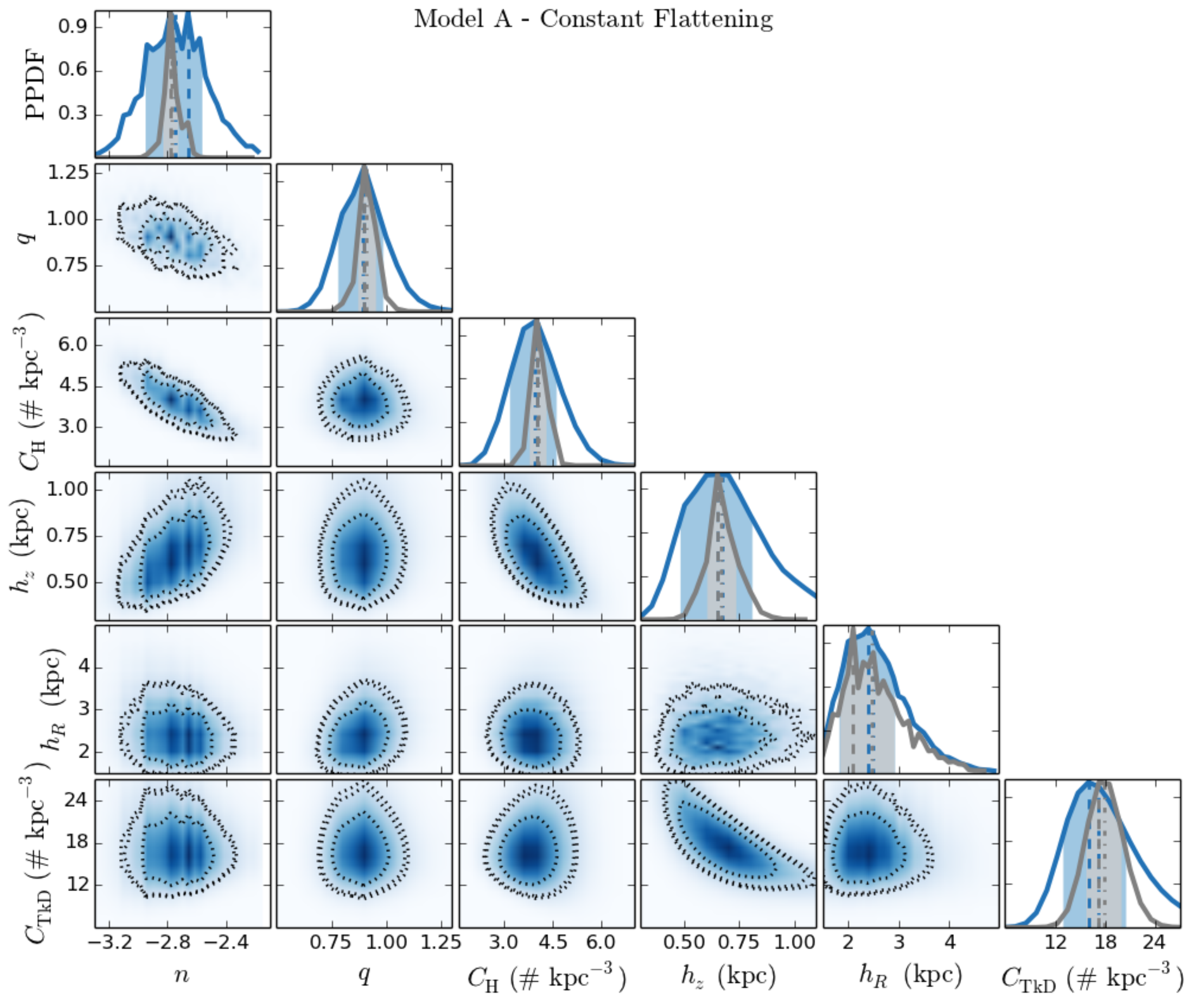}
  \includegraphics[width=1.4\columnwidth]{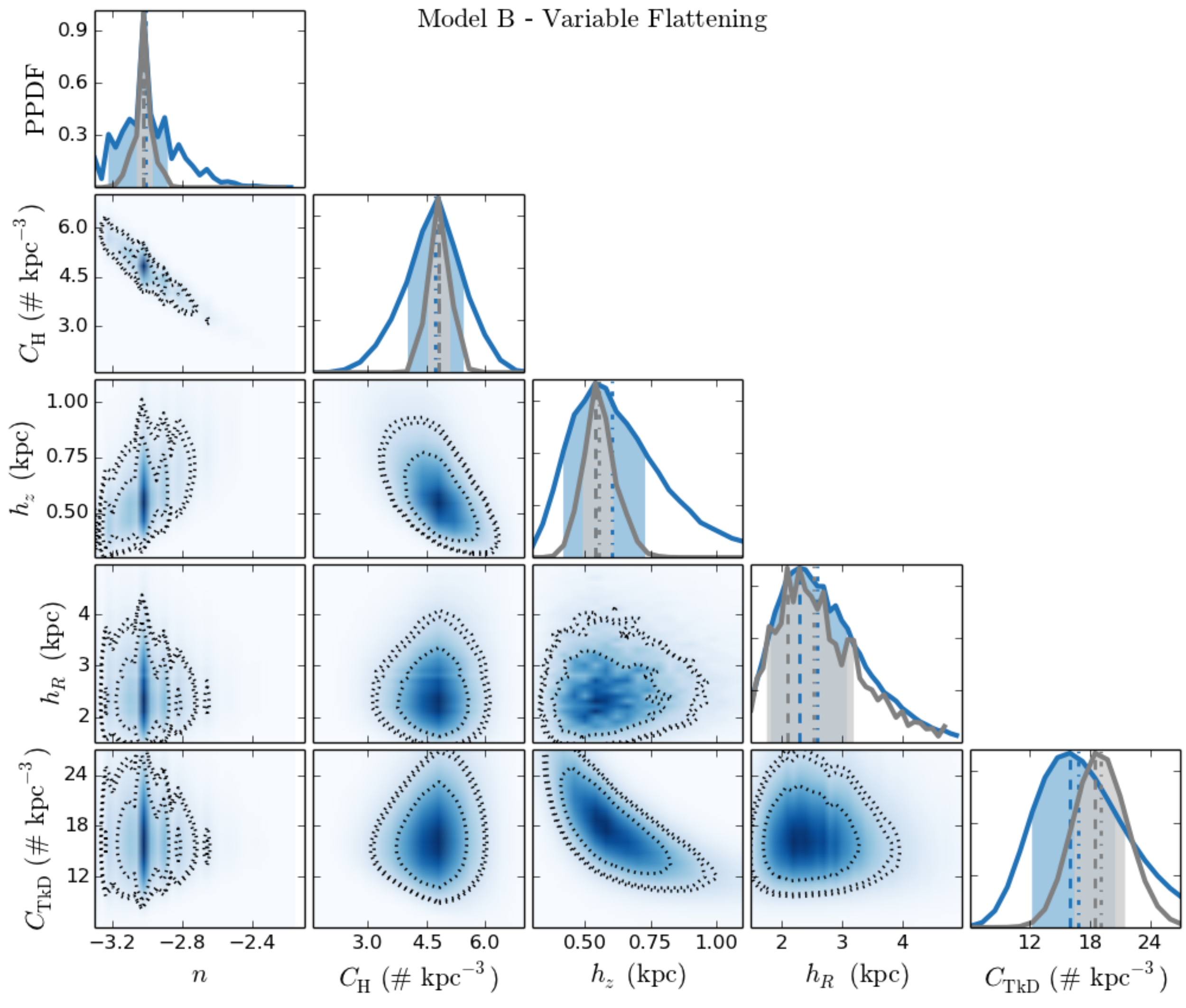}
 \caption{Marginal 2D posterior PDFs for the combined RRLS samples (L95+S10+QUEST), using Model A (top) and model B (bottom). The dotted lines correspond to 1,2 and 3--$\sigma$ confidence regions. \emph{Diagonal}: The dark (blue) and light grey lines show respectively the marginal 1D posterior and the joint 1D posterior, conditional on the best-fitting values for the remaining parameters. The dashed and dotted lines denote each distribution's mode and median respectively, and the shaded regions correspond to 1--$\sigma$ confidence intervals.}
\label{a:sql_all_pdfs}
\end{center}
\end{figure*}

Figure~\ref{a:sql_all_pdfs} shows the full set of 2D marginal posterior PDFs, for halo Models A (top) and B (bottom). In the 2D PDFs, the contours correspond to 1, 2 and 3-$\sigma$ confidence regions\footnote{Taken as confidence intervals around the mode enclosing 68\%, 95\% and 99.5\% of the total probability, as a common analogy to the Gaussian case.}.  The figure also shows, in the diagonal, the 1D marginal posterior PDFs (blue) and joint posterior PDFs (computed at the best-fitting values for the remaining parameters). In these, the dashed and dotted lines correspond respectively to the mode and median of each of the PDFs shown. The shaded regions correspond to the respective 1-$\sigma$ or 68\% confidence intervals, i.e. intervals around the mode containing 68\% of the probability, in 1-D for the marginal case and in 6-D for the joint posterior.

\clearpage

\end{document}

%% file: table_halo_authors.tex
$\mathbf{ -2.78_{-0.05}^{+0.05}}$ & $\mathbf{0.90_{-0.03}^{+0.05}}$  & RRLS & {\bf MV18-A}\\
 $ -2.5 \pm 0.04 $        & $0.79\pm0.02$          & F-MSTOs           & PD15       \\
 $ -2.42         $        & $0.63  $               & RRLS              & S13        \\
 $ -2.3          $        & $0.6   $               & BHB/BSS           & D11        \\
 $ -2.62 \pm 0.04$        & $0.7 \pm 0.02$         & RRLS              & S11        \\
 $ -2.75 \pm 0.07$        & $0.88\pm0.04$          & CMD-fitting       & dJ10       \\
 $ -2.8  \pm 0.07$        & $0.85\pm0.03$          & CMD-fitting       & dJ10       \\
 $\approx-2.5    $        & $\approx 1.0$          & BHB               & DP10       \\        
 $-3.75          $        & $0.983$                & Subdwarfs         & S09        \\
 $ -2.8  \pm 0.2 $        & $0.64\pm0.1 $          & Phot. parallaxes  & J08        \\
 $ -2.5          $        & $0.6  $                & MS phot. parallax & S02        \\
 $ -2.44         $        & $0.76 $                & CMD-fitting       & R00        \\
 $ -2.75         $        & $0.85 $                & CMD-fitting       & R00        \\
\hline
\multicolumn{4}{c}{Constant Flattening - Assumed $q$} \\
\hline
 $ -2.5 \pm 0.4  $        & $0.7$                  & RRLS               & Z14       \\
 $ -2.8 \pm 0.5  $        & $1.0$                  & RRLS               & Z14       \\  
 $ -2.8 \pm 0.4  $        & $0.7$                  & RRLS               & F14       \\
 $ -3.12 \pm 0.2 $        & $1.0$                  & SEGUE calib. stars & C10       \\  
 $ -2.4          $        & $1.0$                  & RRLS               & W09       \\  
 $ -2.43 \pm 0.06$        & $1.0$                  & RRLS               & M08       \\  
 $ -2.7  \pm 0.1 $        & $0.6$                  & RRLS               & V06       \\
 $ -2.5  \pm 0.1 $        & $1.0$                  & RRLS               & V06       \\  
 $ -2.7  \pm 0.2 $        & $1.0$                  & RRLS               & I00       \\  
 $ -3.0  \pm 0.08$        & $1.0$                  & RRLS               & W96       \\  
 $ -3.0          $        & $1.0$                  & RRLS               & S85       \\  
\hline
\multicolumn{4}{c}{Variable Flattening $q(R,z)$}\\
\hline
 $\mathbf{-3.02_{-0.04}^{+0.05}} $ & $\mathbf{0.5 \to 1}$ {\bf (P91\ddag)}& RRLS  & {\bf MV18-B}\\
 $-5.03_{-0.64}^{+0.64}$   & $0.64 \to 0.96$               &  K-giants            &  X17        \\ 
 $-2.96 \pm 0.05$          & $0.57\pm0.02 \to 0.84\pm0.06$ &  RRLS                &  I17        \\ 
 $-4.7 \pm 0.3$            & $0.39\pm0.09 \to 0.81\pm0.05$ &  BHB/K-giants        &  D16a,b     \\ 
 $-4.2 \pm 0.1$            & $0.2\pm0.1 \to 0.8\pm0.03$    &  K-giants            &  X15        \\ 
 $ -3.15 \pm 0.07$         & $0.5 \to 1$ (P91\ddag)        &  RRLS                &  M08        \\ 
 $ -3.1  \pm 0.1 $         & $0.5 \to 1$ (P91\ddag)        &  RRLS                &  V06        \\ 
 $ -3.55 \pm 0.13$         & $\approx 0.65 \to \approx 1.0$& Metal-poor stars$^b$ & CB00        \\ 
 $ -3.47 \pm 0.18$         & $\approx 0.55 \to \approx 1.0$& Metal-poor stars$^c$ & CB00        \\ 
 $ -3.53 \pm 0.08$         & $0.5 \to 1$ (P91\ddag)        &  RRLS                &  W96        \\ 
 $ -3.2  \pm 0.1 $         & $0.5 \to 1$ (P91\ddag)        &  RRLS                &  P91        \\ 

%% file: table_tkd_authors.tex
$\mathbf{2.10_{-0.25}^{+0.82}}$ & $\mathbf{0.65_{-0.05}^{+0.09}}$ & RRLS & {\bf MV18-A} \\  
$\mathbf{2.10_{-0.34}^{+1.10}}$ & $\mathbf{0.54_{-0.05}^{+0.07}}$ & RRLS & {\bf MV18-B} \\  
  $1.9  \pm 0.1 $  & $\cdots $               &  RGB               &   M17       \\    
  $2.2  \pm 0.2 $  & $\cdots $               &  RC                &   B16       \\    
  $2.36 \pm 0.025$ & $0.535 \pm 0.0046$      &  CMD fitting       &   R14       \\    
  $2.43          $ & $0.596           $      &  CMD fitting       &   R14-CA    \\    
  $2.41          $ & $0.549           $      &  CMD fitting       &   R14-CB    \\    
  $2.01 \pm 0.05 $ & $0.655 \pm 0.013 $      &  G-type dwarfs     &   B12       \\    
  $2.0$            & $\cdots$                &  K-giants          &   B11       \\    
  $2.2  \pm 0.35 $ & $0.51  \pm 0.04  $      &  SEGUE calibration stars   &   C10       \\    
  $4.1  \pm 0.4  $ & $0.75  \pm 0.07  $      &  CMD fitting   &   dJ10      \\    
  $3.6  \pm 0.3  $ & $0.9   \pm 0.09  $      &  MS photometric parallaxes   &   J08       \\    
  $3.04 \pm 0.11 $ & $1.06  \pm 0.05  $      &  2MASS RC counts   &   CL05      \\    
  $4.7  \pm 0.2  $ & $0.9   \pm 0.07  $      &  Photometric parallaxes   &   LH03      \\    
  $3.5  \pm 0.5  $ & $0.75            $      &  MS photometric parallaxes   &   S02       \\    
  $ \cdots $       & $0.665 \pm 0.085 $      &  \emph{griz} CMD fitting &   C01       \\    
  $3.7  \pm 0.5  $ & $0.86  \pm 0.2   $      &  \emph{JHK} CMD fitting  &   O01       \\    
  $2.5           $ & $0.8             $      &  CMD fitting       &   R01       \\    
  $2.8  \pm 0.5  $ & $0.76  \pm 0.05  $      &  \emph{UBV} CMD fitting   &   R96       \\    
  $ \cdots $       & $0.66  \pm 0.16  $      &  RRLS   &   K06 \\  
  $ \cdots $       & $0.7^{+0.5}_{-0.3}$     &  RRLS   &   L95 \\  
\hline
\multicolumn{4}{c}{Metal Weak Thick Disc} \\
\hline
  $2.0           $ & $1.36  \pm 0.13  $      & SEGUE calibration stars &   C10       \\
  $3.5  \pm 0.5  $ & $1.03  \pm 0.01  $      & BHB                     &   B08       \\
  $4.5  \pm 0.6  $ & $1.0             $      & Metal-poor stars        &   CB00      \\